\documentclass[journal=langd5,manuscript=article]{achemso}
\usepackage{times}
\usepackage{graphicx}
\usepackage{psfrag}
\usepackage{epsfig}
\usepackage{ae}
\usepackage{amsmath,amssymb}
\usepackage{booktabs}
\usepackage[version=3]{mhchem} 
\usepackage[usenames]{color}
\usepackage{float}
\usepackage{subfigure}

\author{Murilo S. Marques}
\affiliation{Centro das Ciências Exatas e das Tecnologias, Campus Reitor Edgard Santos, Universidade Federal do Oeste da Bahia, Rua Bertioga, 892, CEP 47810-059, Barreiras, Bahia, Brazil}
\alsoaffiliation {Instituto de Física, Universidade Federal do Rio Grande do Sul,  Caixa Postal 15051, CEP 91501-970, Porto Alegre, Rio Grande do Sul, Brazil}

\author{José Rafael Bordin}
\affiliation{Departamento de F\'{i}sica, Instituto de F\'{i}sica e Matem\'{a}tica, 
    Universidade Federal de
    Pelotas. Caixa Postal 354, 96001-970, 
    Pelotas, Brazil.}
    \email{jrbordin@ufpel.edu.br}

\title{Hard core-soft shell particles near repulsive interfaces: interplay between adsorption, aggregation and diffusion}

 \date{\today}

\begin{document}


\begin{abstract}

The behavior of colloidal particles with a hard core and a soft shell has attracted the attention for researchers in the physical-chemistry interface not only due the large number of applications, but due the unique properties of these systems in bulk and at interfaces. The adsorption at the boundary of two phases can provide information about the molecular arrangement. In this way, we perform Langevin Dynamics simulations of polymer-grafted nanoparticles. We employed a recently obtained core-softened potential to analyze the relation between adsorption, structure and dynamic properties of the nanoparticles near a solid repulsive surface. Two cases were considered: flat or structured walls. At low temperatures, a maxima is observed in the adsorption. It is related to a fluid to clusters transition and with a minima in the contact layer diffusion - and is explained by the competition between the scales in the core-softened interaction. Due the long range repulsion, the particles stay at the distance correspondent to this length scale at low densities, and overcome the repulsive barrier as the packing increases, However, increasing the temperature, the gain in kinetic energy allows the colloids to overcome the long range repulsion barrier even at low densities. As consequence, there is no competition and no maxima was observed in the adsorption. 

\end{abstract}
\maketitle

\section{Introduction}
\label{intro}

A multitude of natural process take place at the boundary between two phases while others are started at that interface. This is the case of adsorption \cite{dabrowski2001} - a universal phenomenon in colloidal and surface science \cite{colloid_chapter} whose properties have been studied for a long time \cite{bayliss1911, vold1958, robens2014}. Basically, it's a surface effect which causes affluence (change in concentration) of atoms, molecules or ions at two-phase interfaces \cite{tareq2019} drastically modifying its properties compared to bulk \cite{tang2019}. For spherical colloids and nanoparticles, macromolecules that often show competitive interactions \cite{ciach2013,bartlett2005}, the adsorption and binding at interfaces associated with the various probabilities of aggregation are fundamental for applications in biomedical, environmental, food, and materials engineering \cite{adam2002, adam2012,monika2011, amir2016}. In this context, the computational simulation of colloids at interfaces have evolved enormously and nowadays the structure, dynamics, thermodynamics, phase transitions, and reactivity of colloids in confined environments and interfacial geometries have been widely studied \cite{imperio2007,amir2016, rabe2011,stradner2020}, mainly through attractive and repulsive competing interactions \cite{santos2017}. Such interrelation had as their starting point the famous DVLO theory for charged colloids at 1940's \cite{verwey1948} and the seminal work of Asakura and Oosawa (AO) at 1950's \cite{asakura1954,asakura1958}: these two approaches provide a framework by which attraction and repulsion between colloids may be manipulated \cite{royall2018}. 

Today the current stage of modeling has brought to us a variety of potentials that have being used to portray the competitive interactions between spherical colloids. The competitions, usually between a Short range Attraction and a Long range Repulsion -- the  so-called SALR colloids, arise once distinct conformations compete to rule the suspension behavior~\cite{Shukla08,Somerville20,Cardoso21}. The short range attraction is caused by van der Walls forces or solvent effects~\cite{almarza2014}, while the long range repulsion can be generated by many factors. For instance, it can be caused by electrostatic repulsion in charged colloids and molecules~\cite{Ong15, Campos17}, or by soft shells ad in the case of spherical colloids obtained by PEG aggregates~\cite{colloid1, colloid2,Haddadi20} or even by a polymeric brush, as metallic NP~\cite{Marques20, Lafitte14,Curk14, Nie16,Wang16} decorated with polymers and star polymers~\cite{Bos19}.

From experimental~\cite{colloid1, colloid2} and computational~\cite{Marques20,Lafitte14} works, is well known that the SALR NP and colloids effective interaction can be depicted by core-softened potentials. In this way, many works have been devoted to study the behavior of competitive colloidal systems in bulk solutions~\cite{Ja98,Saija09, Malescio11, Prestipino10,Prestipino12, Cos13}. There are a plenty of works exploring the behavior of confined SALR systems in recent times: Almarza et al have inspected the template-assisted pattern formation in monolayers of particles by Monte Carlo simulations in a lattice gas generic model\cite{almarza2014, almarza2016}; Litniewski and Ciach have analyzed the general features of adsorption phenomena in dilute systems with particles self-assembling into small clusters \cite{ciach2019jcp, ciachsoft2019}; more recently, Panagiotopoulos et al have scrutinized properties of lamellar structures formed by an SALR fluid in equilibrium and non-equilibrium conditions by machine learning \cite{grego2020}; Bilnadau et al investigated the cluster formation effect on adsorption phenomena they have laid down deviation in the shape of the adsorption isotherm in comparison with simple fluids \cite{ciach2020}. Though, according to the authors' knowledge, no relationship between aggregation, adsorption and dynamic management was listed in these SALR colloids at interfaces. In this work, we have looked over the behavior of a SALR system confined by two types of plates: rough and flat, and we've explored the distinctness of adsorption isotherms, lateral structuring (by means of the lateral radial distribution function) and the dynamic behavior of the system in order to answer the question: how the surface smoothness influence the adsorption in systems shaped by competitive interactions?

The answer to this inquiry is organized as follows. In Sec. II, we provide details of model and simulation details,  while results and theoretical approach are presented and discussed in Sec. III. Conclusions follow in Sec. IV.

\section{The Model and Simulation Details}
\label{Model}

    \setcounter{subfigure}{0}
    \begin{figure}[ht]
        \centering
        \subfigure[]{\includegraphics[width=0.45\textwidth]{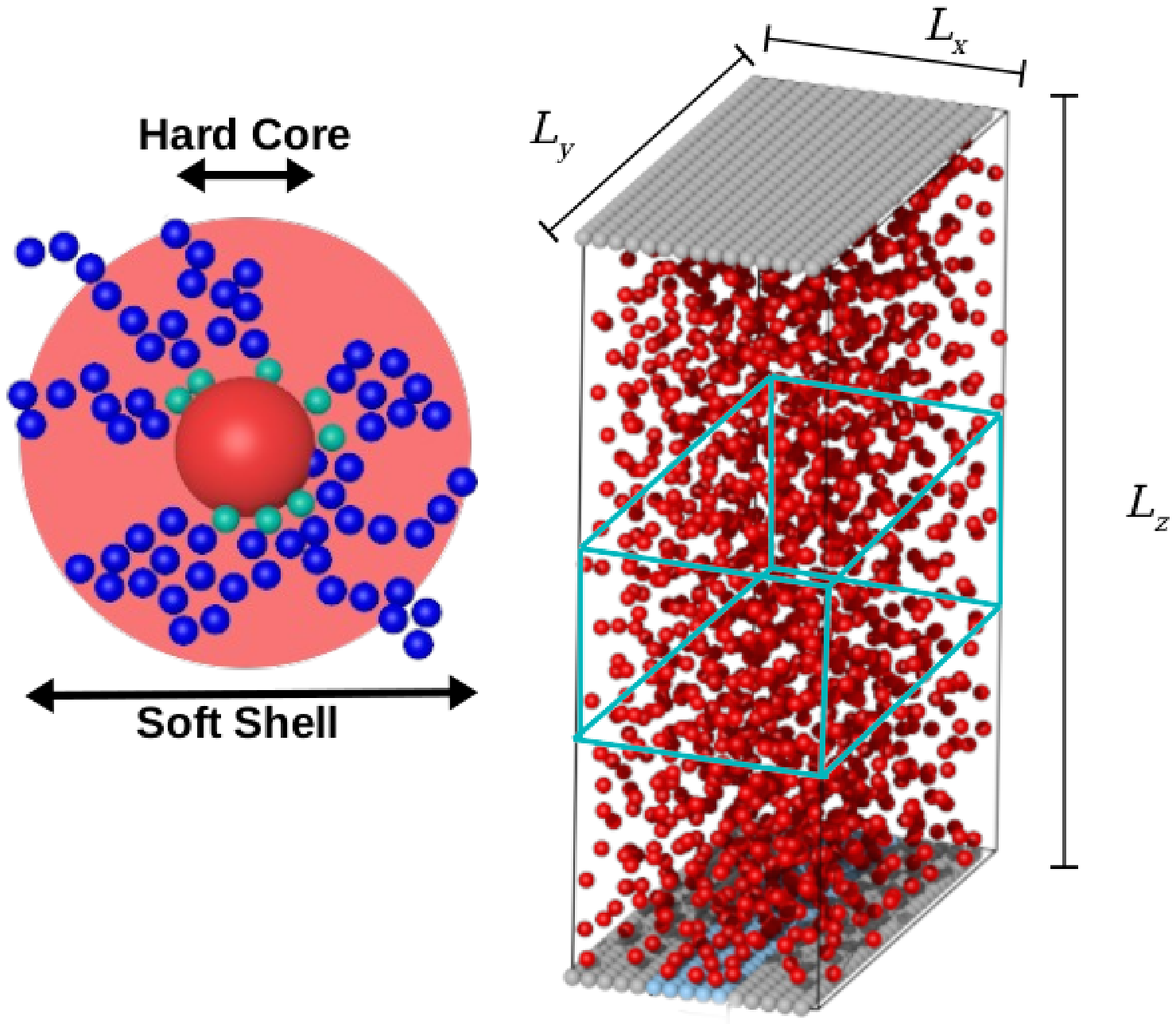}}
        \subfigure[]{\includegraphics[width=0.45\textwidth]{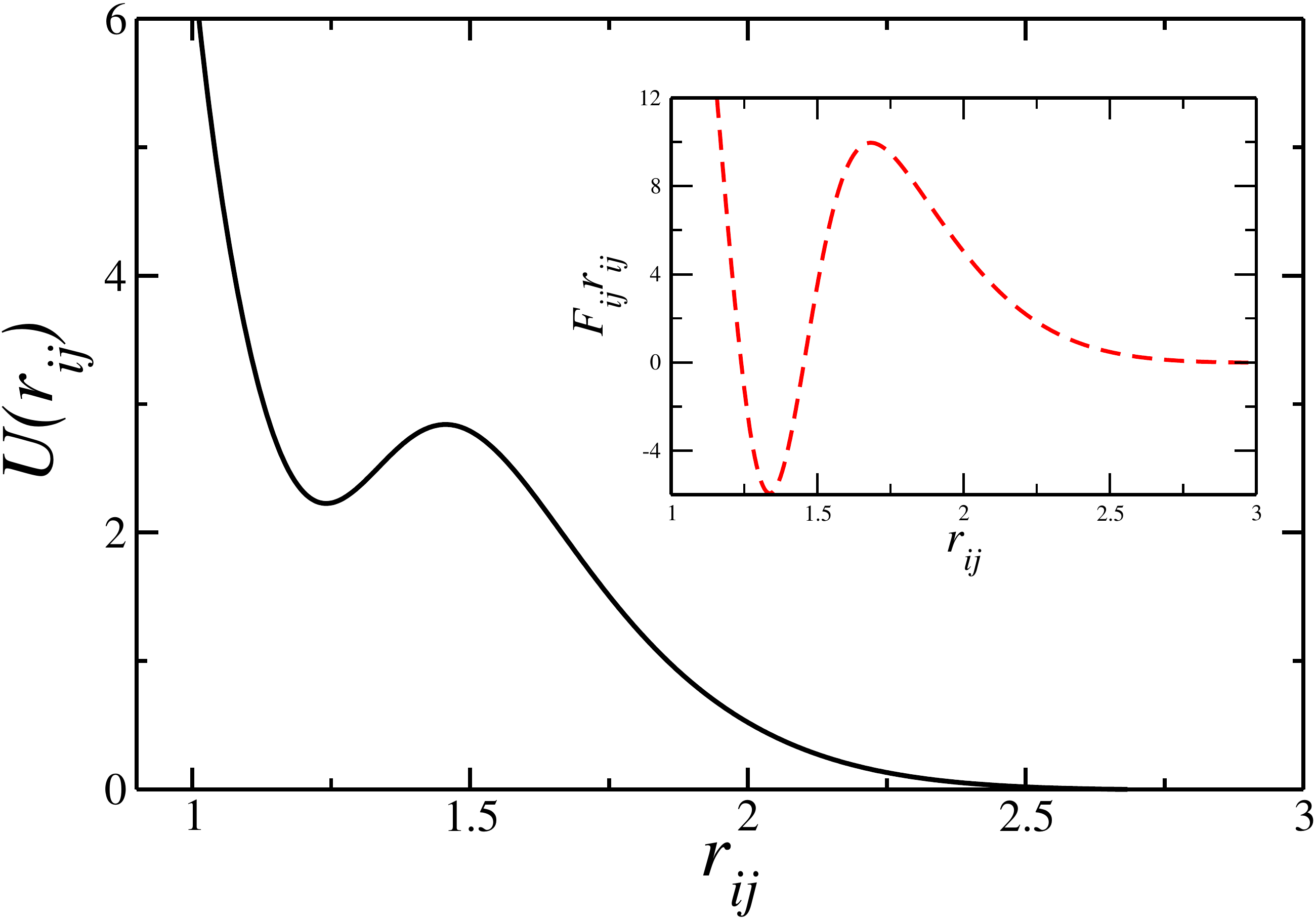}}
        \caption{(a) Schematic depiction of the polymer-grafted NP and simulations box. In the NP, the blue monomers stand for the polymers, the green is the monomer connected to the hard-core red bead. The simulation box with size $L_x \times L_y \times L_z$ in the $x$, $y$ and $z$ directions has two walls in the $z$-extremes. The turquoise lines represents the density control volume region. (b) Effective potential $U(r_{ij})$ employed to model the colloids as obtained in our recent work~\cite{Marques20}. The inset is the product of the pair force $F_{ij} = -dU(r_{ij})/dr_{ij}$ and the distance $r_{ij}$ as function of the distance. }
        \label{fig1}
    \end{figure}

The fluid consists of spherical nanoparticles (NP) with a hard core and a soft corona, as schematically depicted in the figure~\ref{fig1}(a).  The effective core-softened interaction potential, obtained in a recent work by Marques and co-authors~\cite{Marques20}, is composed by a short-range attractive Lennard Jones potential and three Gaussian terms, each one centered in $c_j$, with depth $h_j$ and width $w_j$:
\begin{equation}
U(r_{ij})=4\epsilon\left [\left (\frac{\sigma}{r_{ij}}  \right )^{12} - \left (\frac{\sigma}{r_{ij}}  
\right )^{6}  \right ]+ \sum_{j=1}^{3}h_jexp\left [ -\left ( \frac{r_{ij}-c_j}{w_j} \right )^2 \right ], 
\label{pot}
\end{equation}
\noindent Here, $r_{ij} = |\vec r_i - \vec r_j|$ is the distance between two particles $i$ and $j$. The Gaussian parameters are given in table~\ref{table1}. With these parameters, the interaction potential has a shoulder-like shape and two characteristic interaction length scales~\cite{Evy2011,Formin2011,Ney2009,jonatan2010}, as shown in the figure~\ref{fig1}(b). A closer length scale, located near $1.2\sigma$, that stands for the hard-core, and a further second scale $r_{ij} \approx 2.2\sigma$ from the soft-corona. Between them there is a entropic barrier - the scales and the barrier are clear in the inset of the figure~\ref{fig1}(b).

\begin{table}
\centering
\setlength{\tabcolsep}{12pt}
\begin{tabular}{c | c }
\hline
Parameter & Value  \\ 
\hline 
$h_1$ & 3.50803  \\ 
$c_1$ & 1.05317  \\ 
$w_1$ & 0.0887196 \\ 
$h_2$ & 3.2397  \\ 
$c_2$ & 1.37689  \\ 
$w_2$ & 0.468399  \\
$h_3$ & -3.8685 \\ 
$c_3$ & 1.1684   \\ 
$w_3$ & 0.2400\\ 
\hline 
\end{tabular}
\caption{Interacting potential parameters in reduced units.}
\label{table1}
\end{table}

Langevin Dynamics simulations were performed with a time step of $\delta t = 0.01$ and $\gamma = 1.0$ using the ESPResSo package~\cite{espresso1, espresso2}. The simulation box has dimensions $L_x = 15\sigma$, $L_y = 30\sigma$ and $L_z = 45\sigma$. The adsorption surfaces are confining plates placed in the limits of the $z$-direction. $V_F = [L_x\times L_y\times (L_z-2.0)]$ is the available volume for the particles considering the excluded volume from the wall particles. Periodic boundary conditions were applied in the $x$ and $y$-directions. 

The confining plates can be flat or rough. In the rough case they are modeled as spherical particles with diameter $\sigma$ distributed in a square lattice and fixed in space. The interaction between the wall and fluid particles is given by the purely repulsive Weeks-Chandler-Andersen (WCA) potential. The WCA interaction is a LJ interaction -- first therm in Equation~\ref{pot} -- cut at $r_{ij} = 2^{\frac{1}{6}}\sigma$ and shifted by $\epsilon$. In the flat case the wall was considered as a plane that repels the fluid by the projection of the WCA potential in the $z$-direction - interacting as a structureless flat surface.

Simulations were performed with different bulk number density, ranging from $\rho_0 = N_0/V_F = 0.025\sigma^3$ to $\rho_0 = 0.200\sigma^3$ and temperatures from $T = 0.1 \epsilon/k_B$ to $T = 0.90\epsilon/k_B$. The initial number of particles in the system, obtained from $N_0 = \rho_0 V_F$, were initially randomly distributed in the simulation box. However, considering that particles will adsorb and get structured near to the wall, we create a Control Volume (CV) at the center of the simulaton box to control the bulk density $\rho_0$, as depicted in the figure~\ref{fig1}(a). The control volume has dimensions $L_x \times L_y \times L_z/3$. Unlike previous works, were a Grand Canonical Monte Carlo simulation was used to keep the chemical potential fixed in a CV~\cite{Bordin12, Bordin13, Bordin14}, here we adopted a simpler and faster approach: we control the density in the CV at every 500 time steps during the thermalization steps. If it deviated more than 2\% from the initial value $\rho_0$ we insert/remove particles to restore the desired density. We check for overlaps and new particles are inserted with a initial velocity obtained from a Gaussian distribution at the proper thermal energy. We observe that after $5\times10^5$ thermalization steps the density in the CV do not deviates more than 2\% from the mean value and, therefore, no more insertion/deletion moves are performed. The thermalization steps are followed by $5\times10^5$ steps to equilibrate the system. Finally, we run $2\times10^6$ steps for the results production stage. To ensure that the system was thermalized, the pressure, kinetic and potential energy  were analyzed as function of time. The velocity-verlet algorithm was employed to integrate the equations of motion. 5 independent simulations (distinct initial random positions and velocities) were performed and here we present the average of this 5 results -- the errors bars are smaller than the points in the results shown in Section~\ref{Results}.
 
 The Gibbs adsorption isotherms were evaluated using
 
 \begin{equation}
 \Gamma(\rho_0) = \int_0^{\infty} [\rho(z) - \rho_0] dz,
 \end{equation}
 where $\rho(z)$ is the density profile along the $z$-direction and $\rho_0$ the density in the CV. The lateral dynamics in the non-confined plane was analyzed by the relation between the lateral mean square displacement (LMSD) and time, namely
\begin{equation}
\left \langle [r(t)-r(t_0)]^2 \right \rangle = \left \langle \Delta r^2(t) \right \rangle,
\end{equation}
where $r(t_0) = (x^2(t_0) + y^2(t_0))^{1/2}$ and $r(t) = (x^2(t) + y^2(t))^{1/2}$ denote the coordinate of  the particle at a time $t_0$ and at a later time t, respectively. The LMSD is related to the diffusion coefficient $D$ by\cite{allen2017}
\begin{equation}
D = \lim_{t\rightarrow \infty } \frac{\left \langle \Delta r^2(t) \right \rangle}{4t}.
\end{equation}
The structure of the fluid was analyzed using the lateral radial distribution function (RDF) $g_{||}(r_{ij})$~\cite{Bordin15},
\begin{equation}
 g_{||}(r_{ij}) \equiv \frac{1}{\rho^2 V}\sum_{i\neq j} \delta(r - r_{ij}) [\theta(|z_i-z_j|) -  \theta(|z_i=z_j|-\delta z]
\end{equation}
 where the Heaviside function $\theta(x)$ restricts the sum of particles pairs in a slab of thickness $\delta z = 1.5\sigma$.
 
 The clustering was analyzed based in the inter particle bonding~\cite{Toledano09, Bordin19, Bordin18b}. Two particles in a layer belong to the same cluster if the distance between two of them is shorter than a cutoff 1.3 - a value slightly bigger than the first lenght scale. $n_c$ is the number of particles in each cluster, and $P(n_c)$ the probability to find a cluster with size $n_c$.

In this work all the quantities are computed and presented in the standard Lennard Jones (LJ) reduced units~\cite{allen2017},
\begin{equation}
\label{red1}
r^*\equiv \frac{r}{\sigma}\;,\quad \rho^{*}\equiv \rho \sigma^{3}\;, \quad 
\mbox{and}\quad t^* \equiv t\left(\frac{\epsilon}{m\sigma^2}\right)^{1/2}\;,
\end{equation}
\noindent for distance, density of particles and time , respectively, and
\begin{equation}
\label{rad2}
p^*\equiv \frac{p \sigma^{3}}{\epsilon} \quad \mbox{and}\quad 
T^{*}\equiv \frac{k_{B}T}{\epsilon}
\end{equation}
\noindent for the pressure and temperature, respectively, where $\sigma = 1.4$ nm, $\epsilon_{core}/k_B = 10179$ K, with $k_B$ the Boltzmann constant, are the distance and energy parameters as previously works~\cite{Marques20, Lafitte14}. $m$ is the mass of a single NP. Since all physical quantities are defined in reduced LJ units, the $^*$ will be omitted, in order to simplify the discussion.

\section{Results and discussion}
\label{Results}

    \setcounter{subfigure}{0}
    \begin{figure}[ht!]
        \centering
        \subfigure[]{\includegraphics[width=6cm,height=3cm]{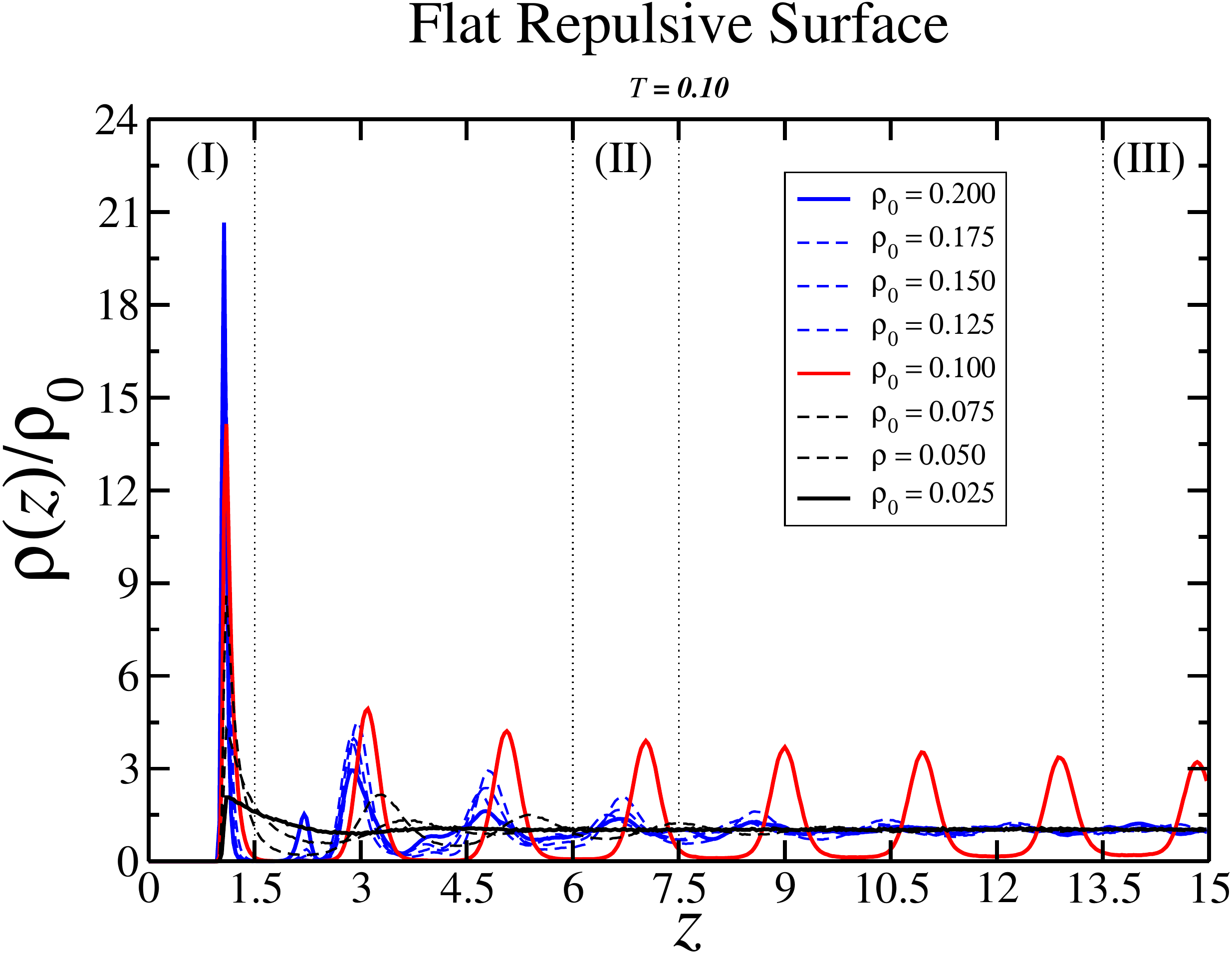}}
        \subfigure[]{\includegraphics[width=6cm,height=3cm]{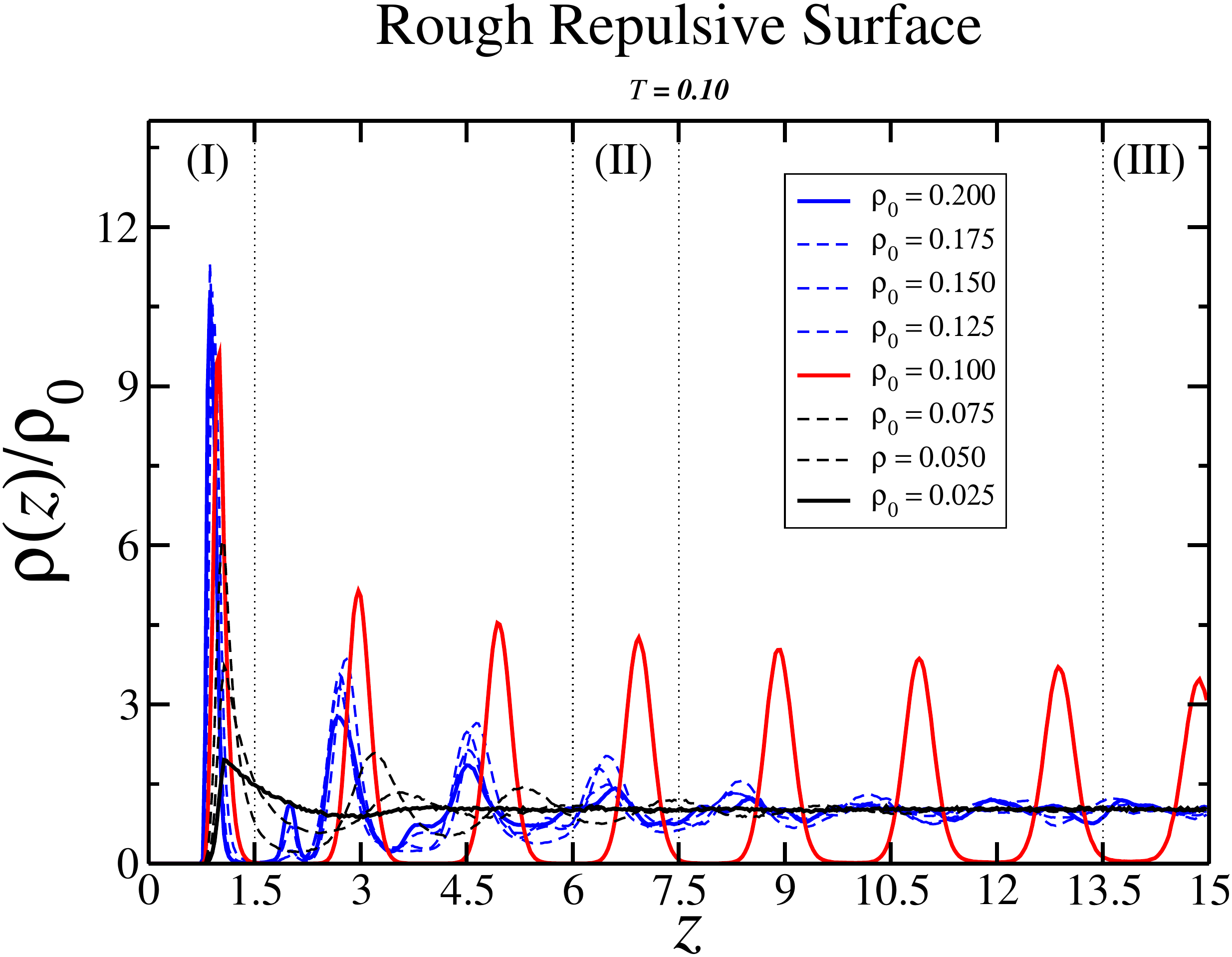}}
        \subfigure[]{\includegraphics[width=6cm,height=3cm]{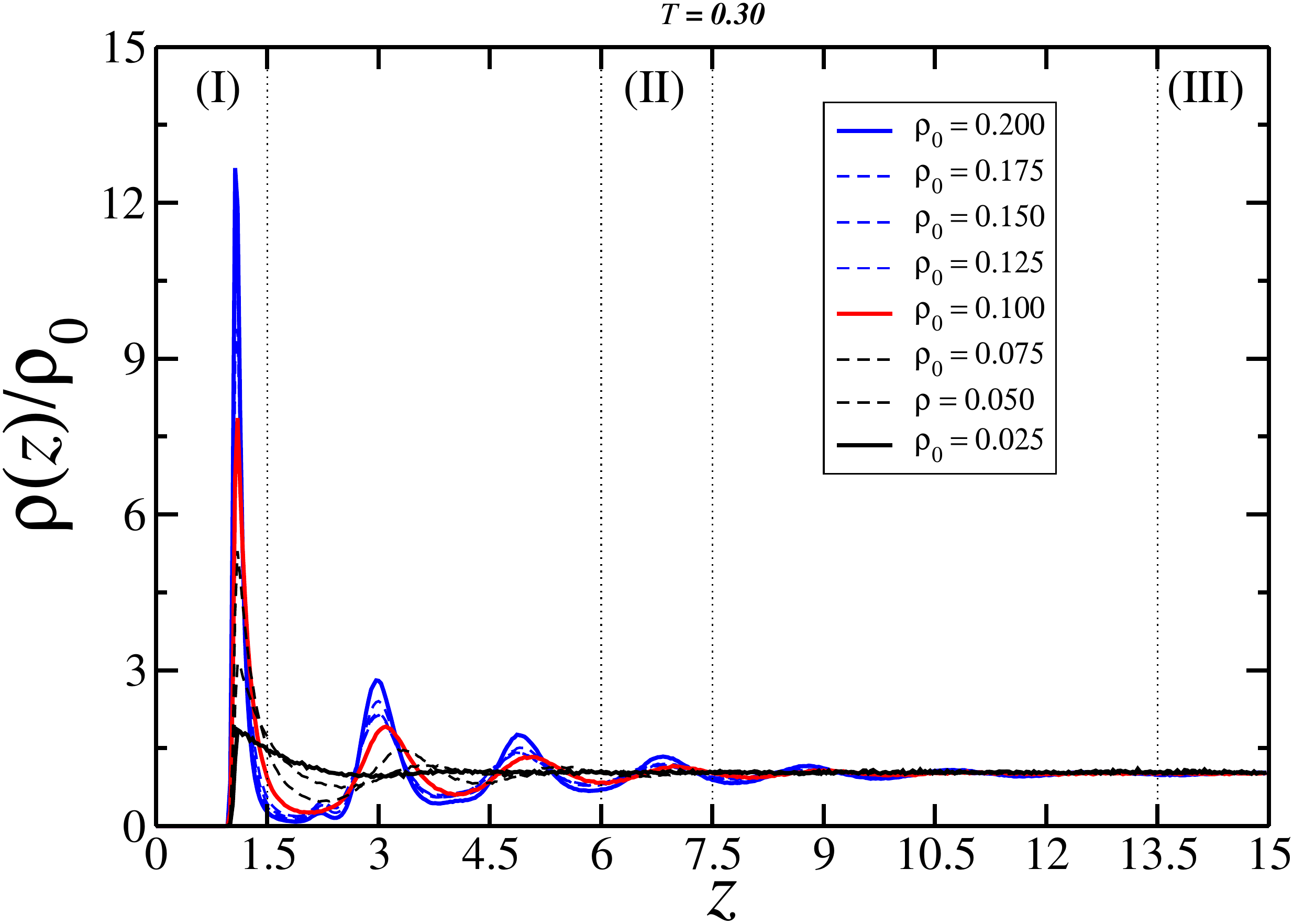}}
        \subfigure[]{\includegraphics[width=6cm,height=3cm]{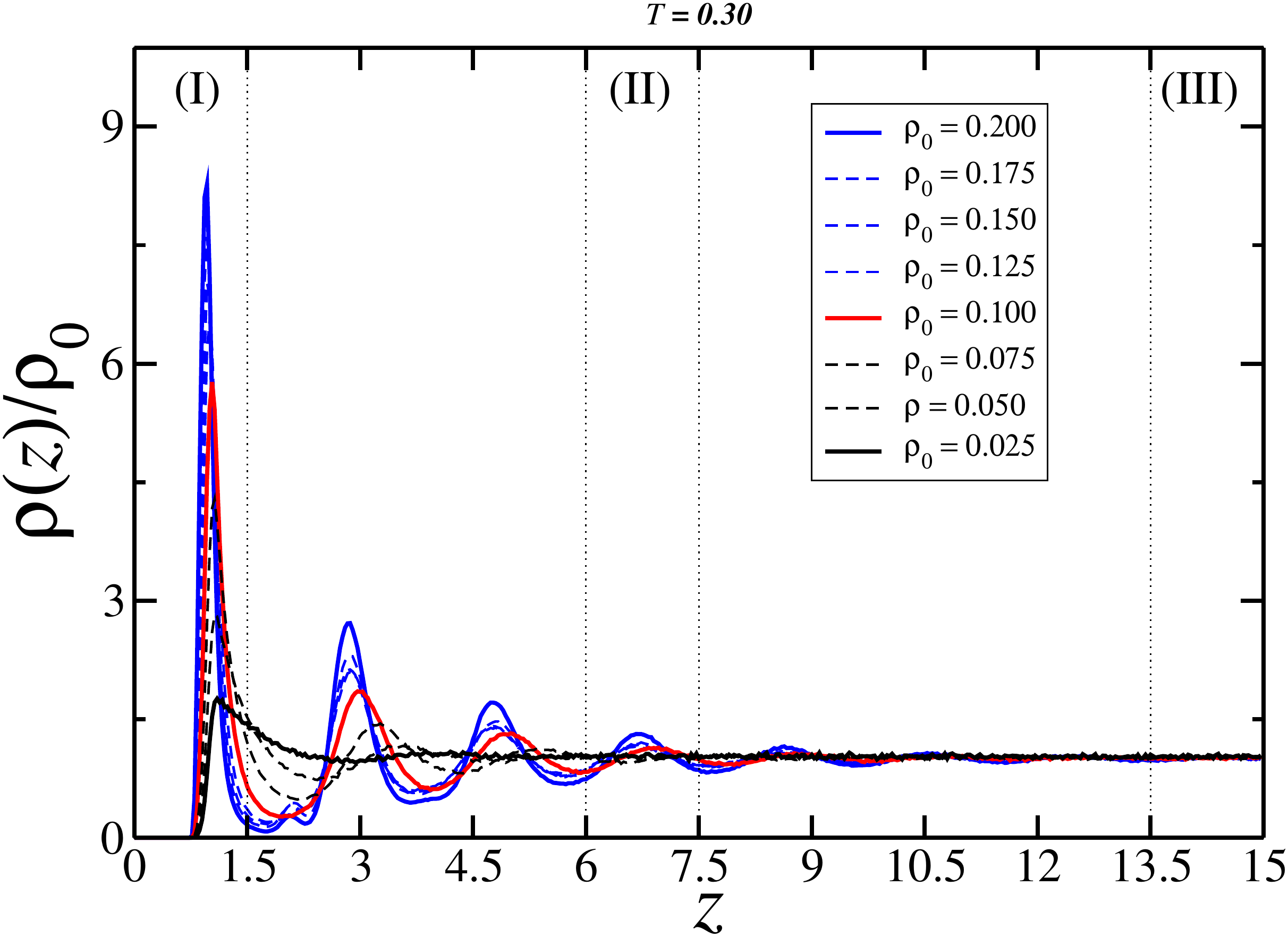}}
        \subfigure[]{\includegraphics[width=6cm,height=3cm]{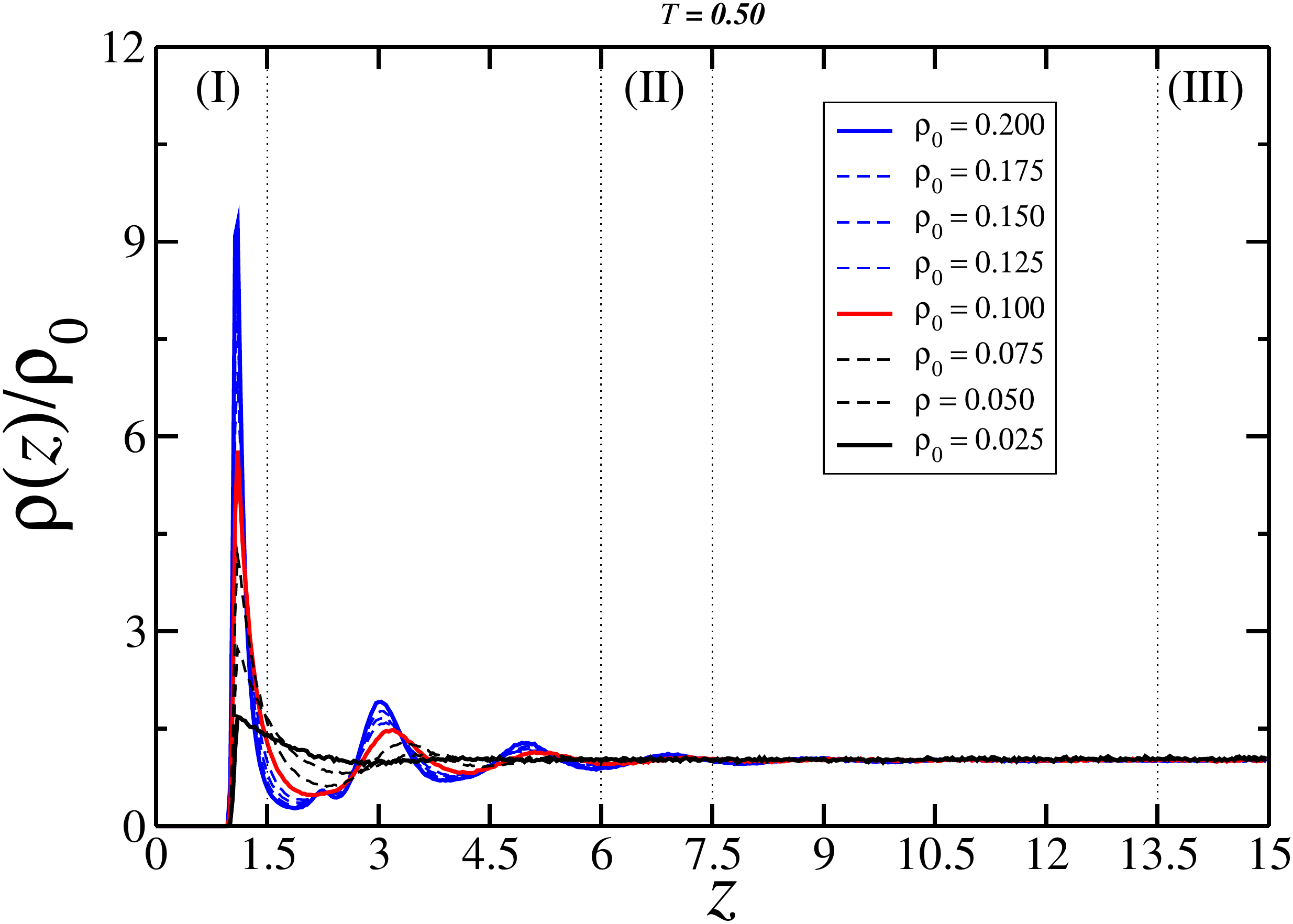}}
        \subfigure[]{\includegraphics[width=6cm,height=3cm]{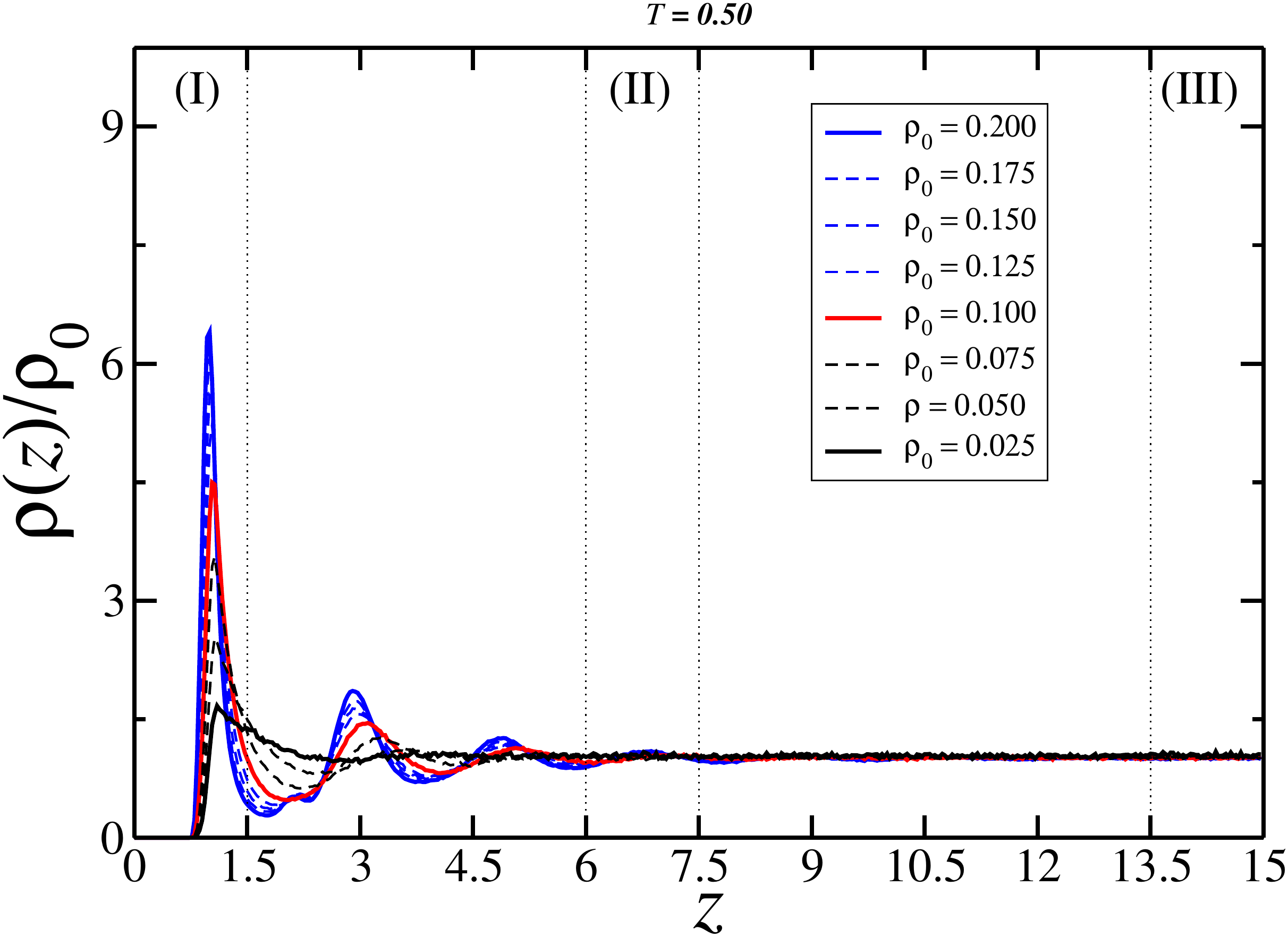}}
        \subfigure[]{\includegraphics[width=6cm,height=3cm]{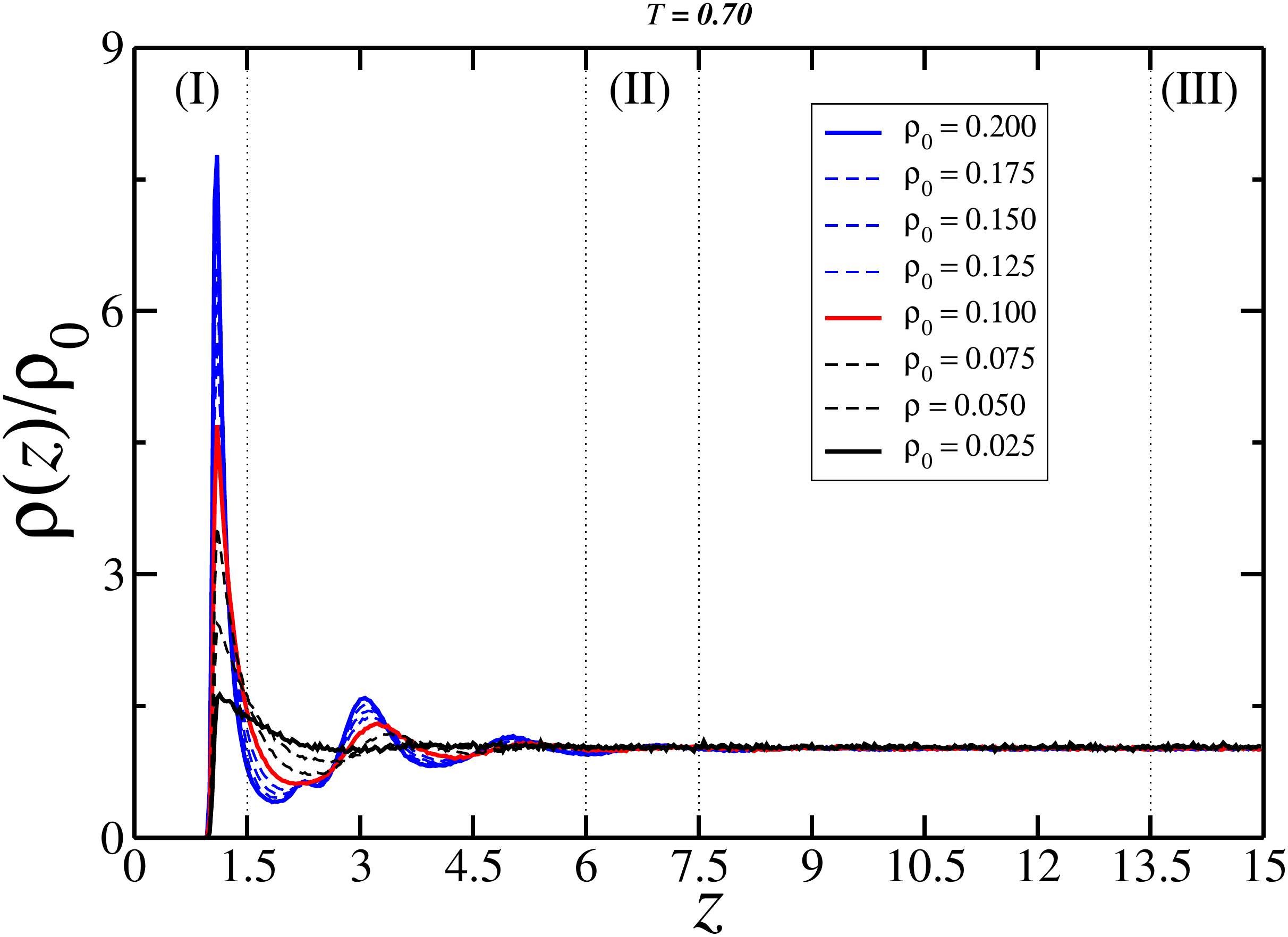}}
        \subfigure[]{\includegraphics[width=6cm,height=3cm]{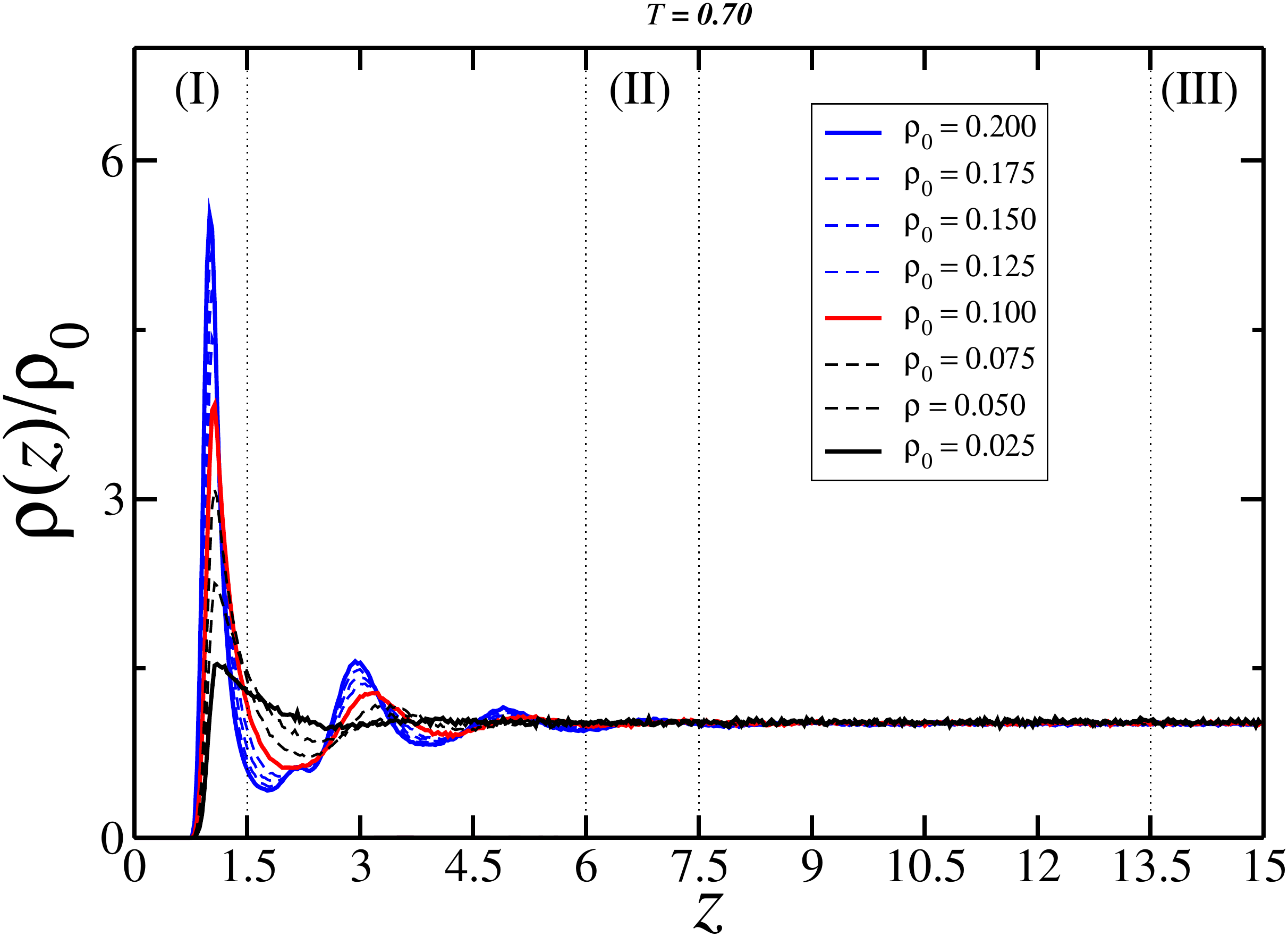}}
        \subfigure[]{\includegraphics[width=6cm,height=3cm]{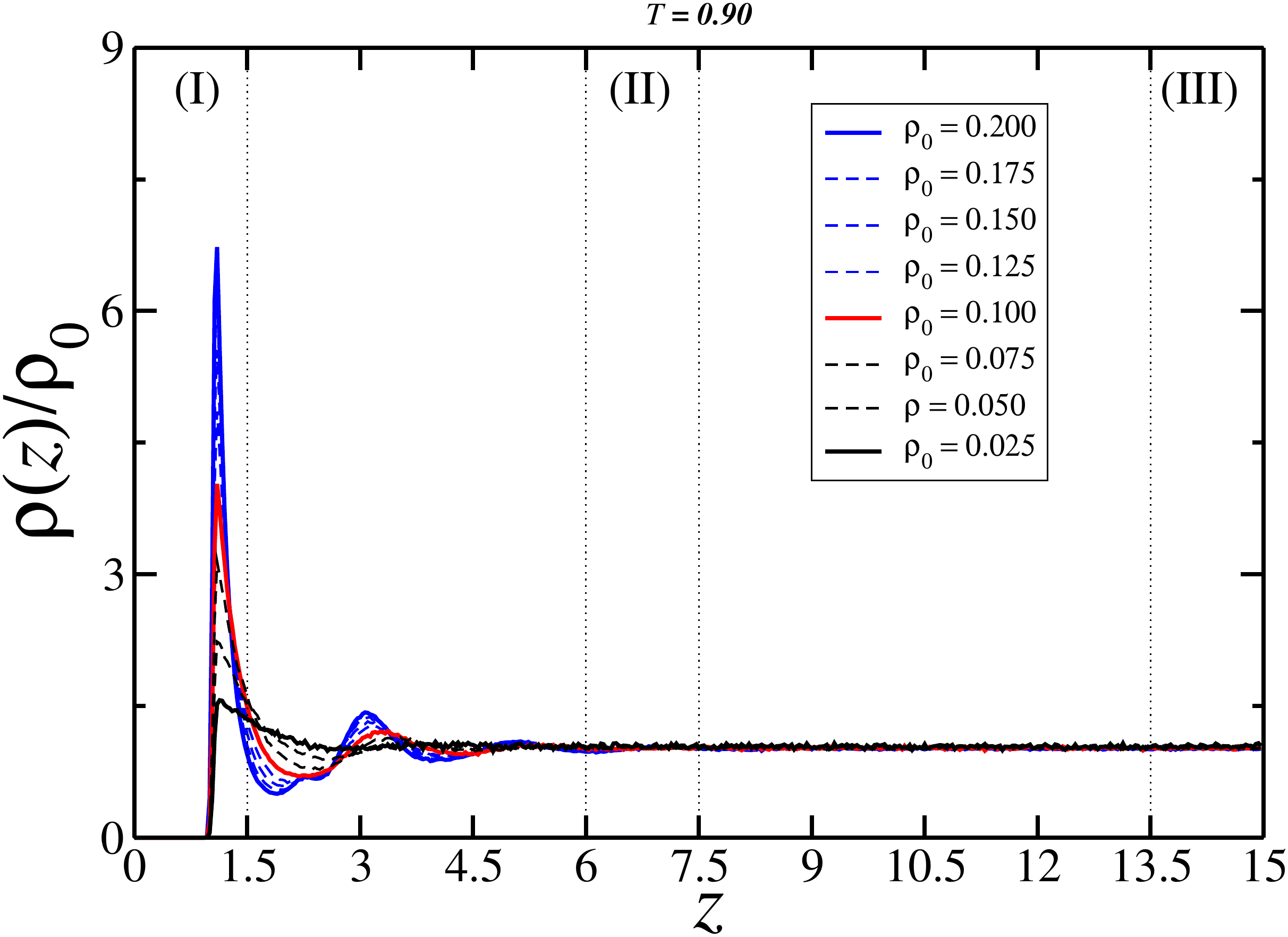}}
        \subfigure[]{\includegraphics[width=6cm,height=3cm]{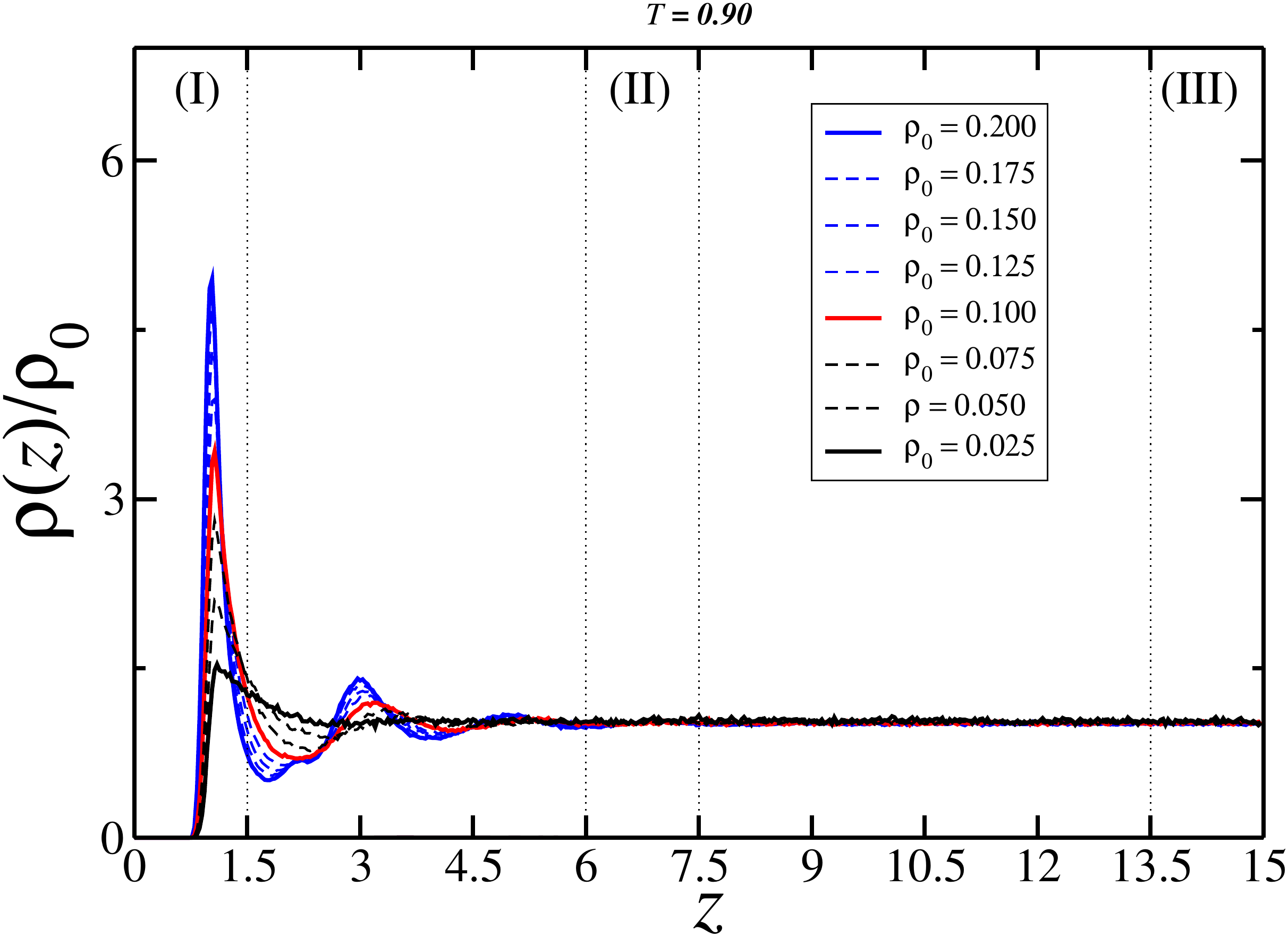}}
         \caption{Ramp-like colloids density profile along the $z$-direction near repulsive walls.}
        \label{fig2}
    \end{figure}

To analyze the layering near the surfaces we have evaluated the density profile along the confining direction. In figure~\ref{fig2} we show the densities profiles $\rho(z)$ normalized by the bulk density $\rho_0$. To analyze the properties of distinct fluid layers we have sliced the system in slabs with thickness $\delta z = 1.5$. Then, we select three slabs: the contact slab (I), that ranges from $z = 0.0$ to $z =1.5$, one slab in the bulk region (III), from $z = 13.5$ to $z = 15.0$, and a intermediate slab (II) from $z = 6.0$ to $z = 7.5$, as indicated by the vertical dotted lines in the figure~\ref{fig2}.

As expected, the layering is influenced by the temperature - and so adsorption isotherm with distinct behaviors. System at the smaller temperature have a layering that can spam to the entire simulation box - as we can see for $T = 0.10$ and $\rho = 0.100$ for both surfaces. For higher temperature there is only the creation of a contact layers and a second layer at high densities. Is also clear that the surface roughness do not affect drastically the layering, but make small changes in the occupancy in each layer. Analyzing only the contact layer is clear that, for all combinations of densities and temperatures, the shape of this peak in the $\rho(z)$ curve are distinct for each wall: in flat surfaces the adsorption peak are higher and thinner, while for the structured surfaces they are lower and thicker.

    \setcounter{subfigure}{0}
    \begin{figure}[ht!]
        \centering
    \subfigure[]{\includegraphics[width=0.4\textwidth]{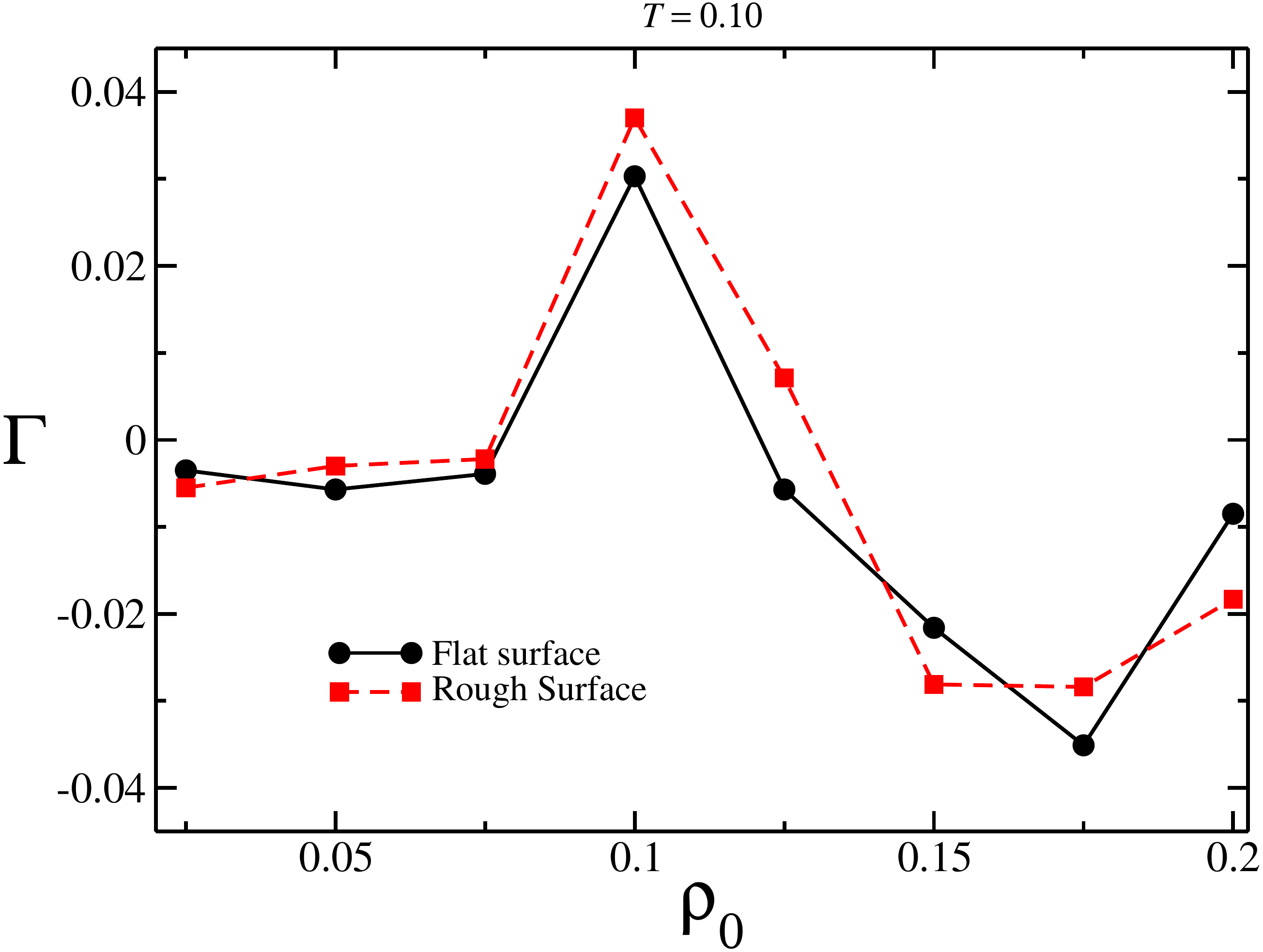}}
    \subfigure[]{\includegraphics[width=0.4\textwidth]{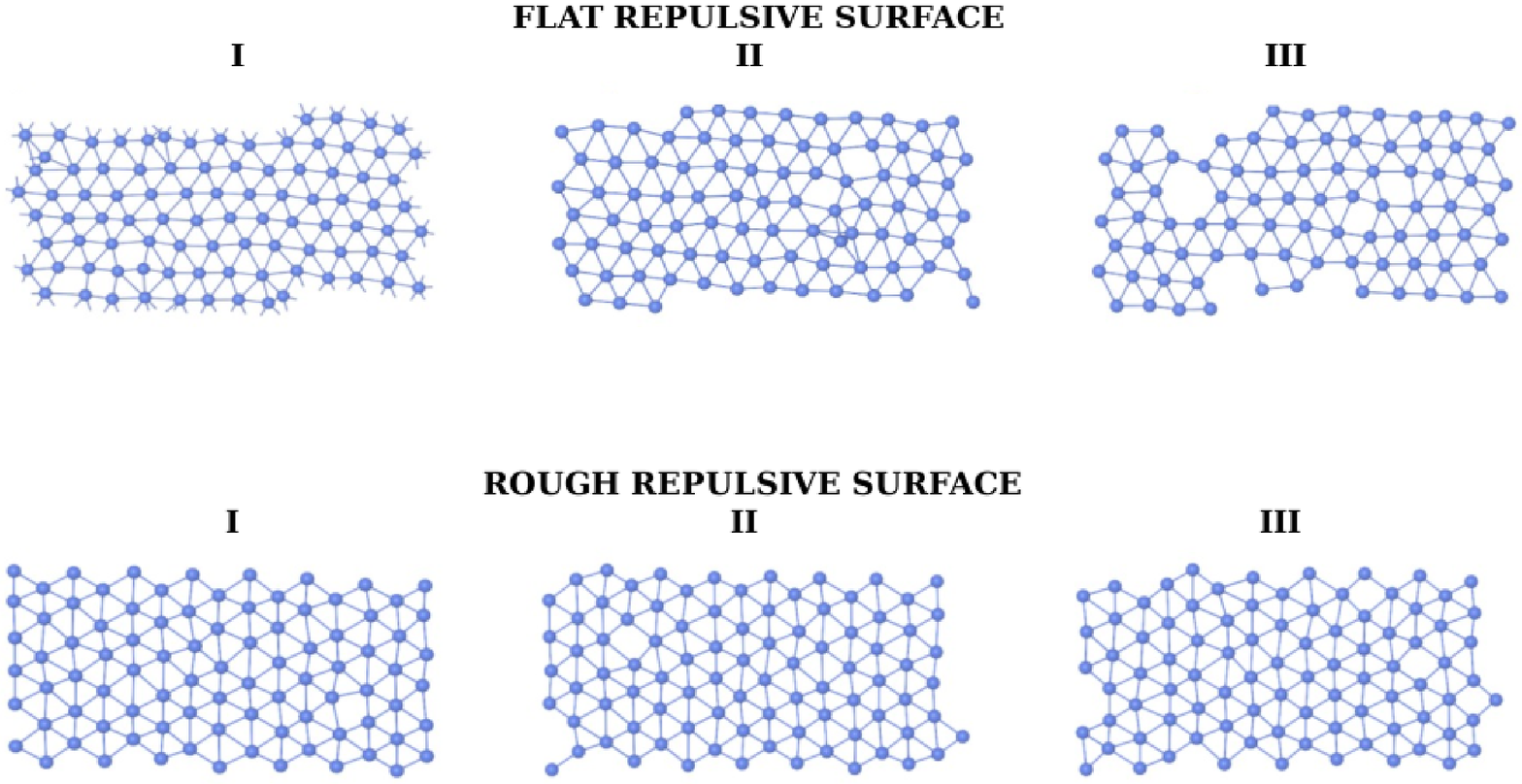}}
    \subfigure[]{\includegraphics[width=0.4\textwidth]{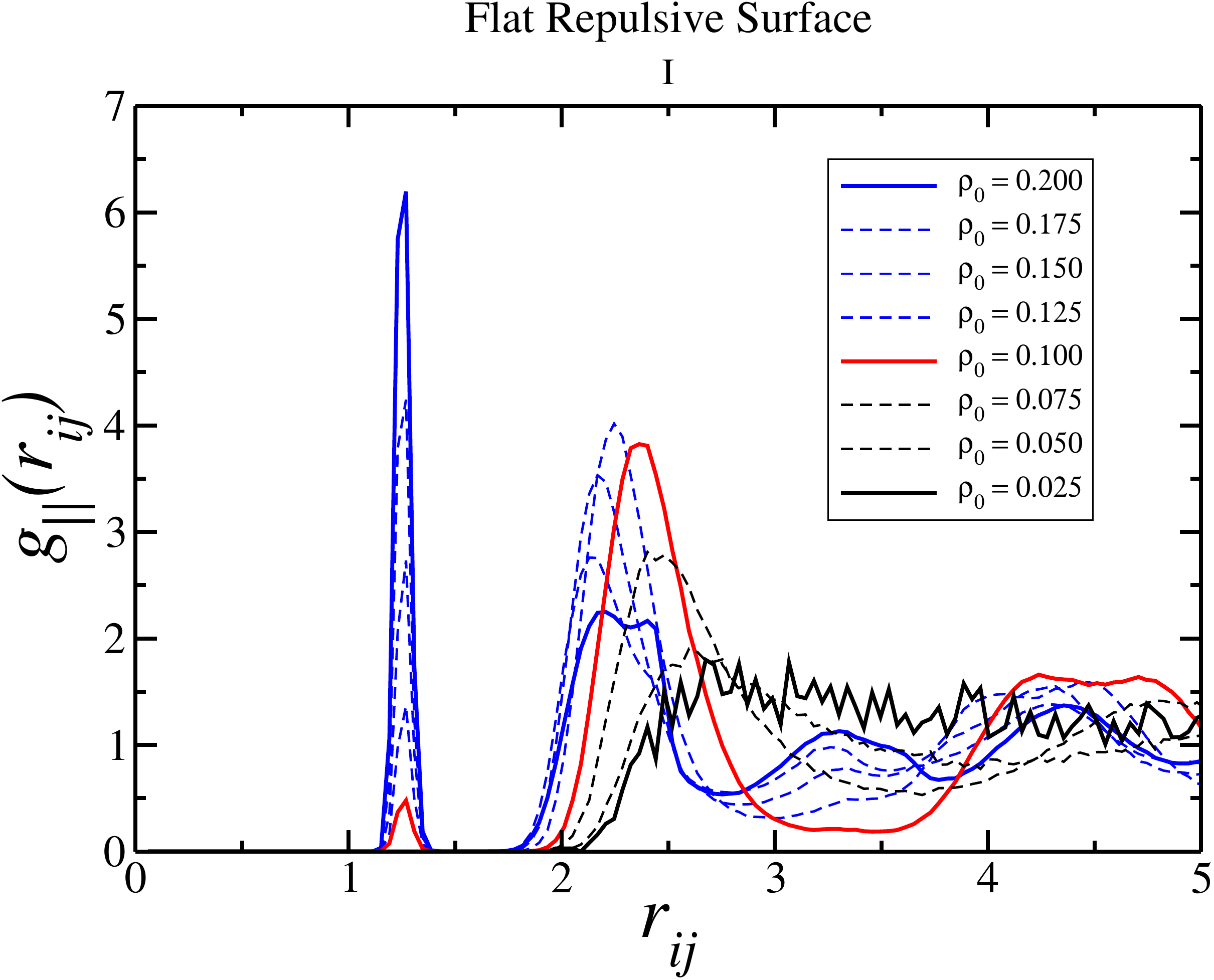}}
    \subfigure[]{\includegraphics[width=0.4\textwidth]{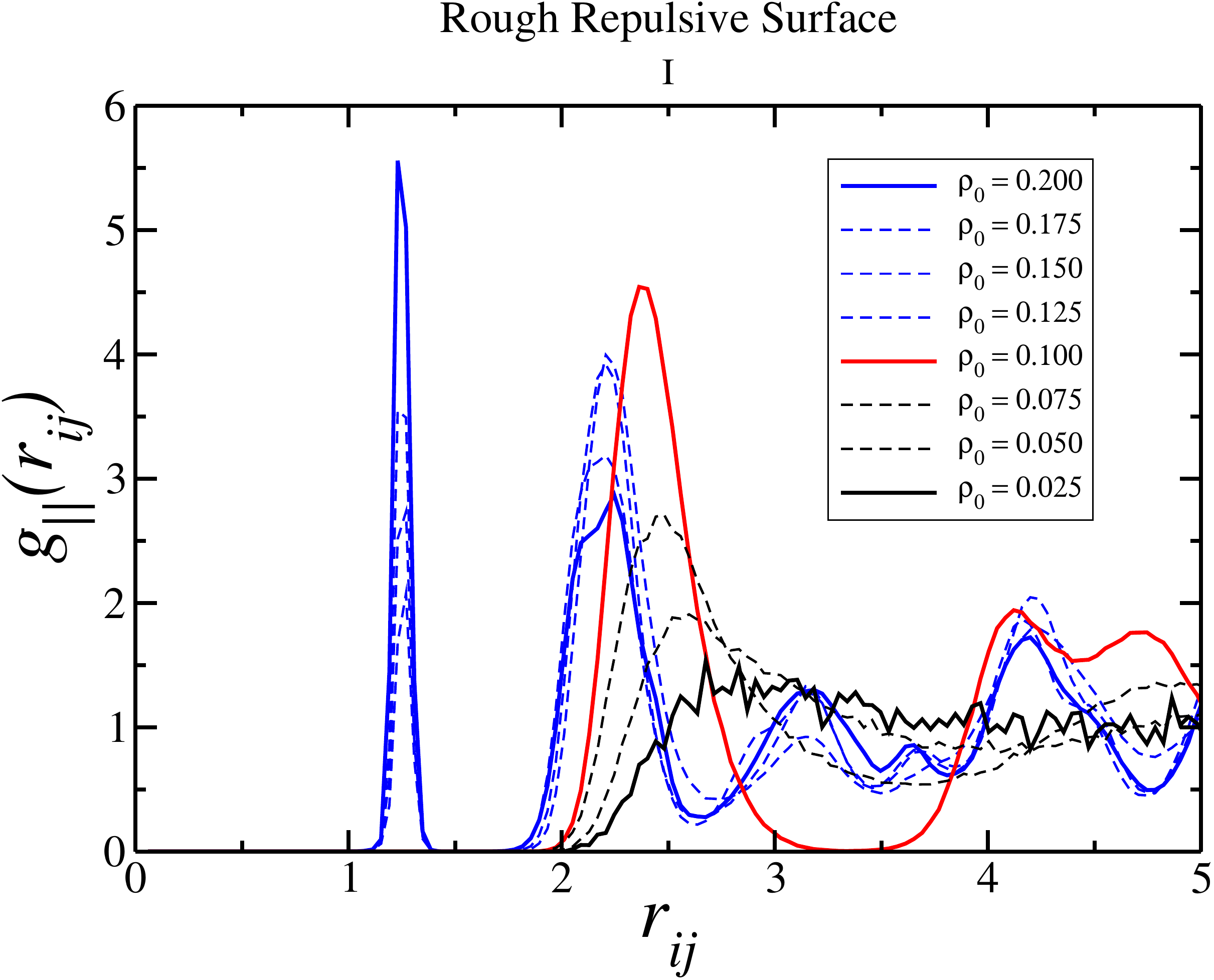}}
    \subfigure[]{\includegraphics[width=0.4\textwidth]{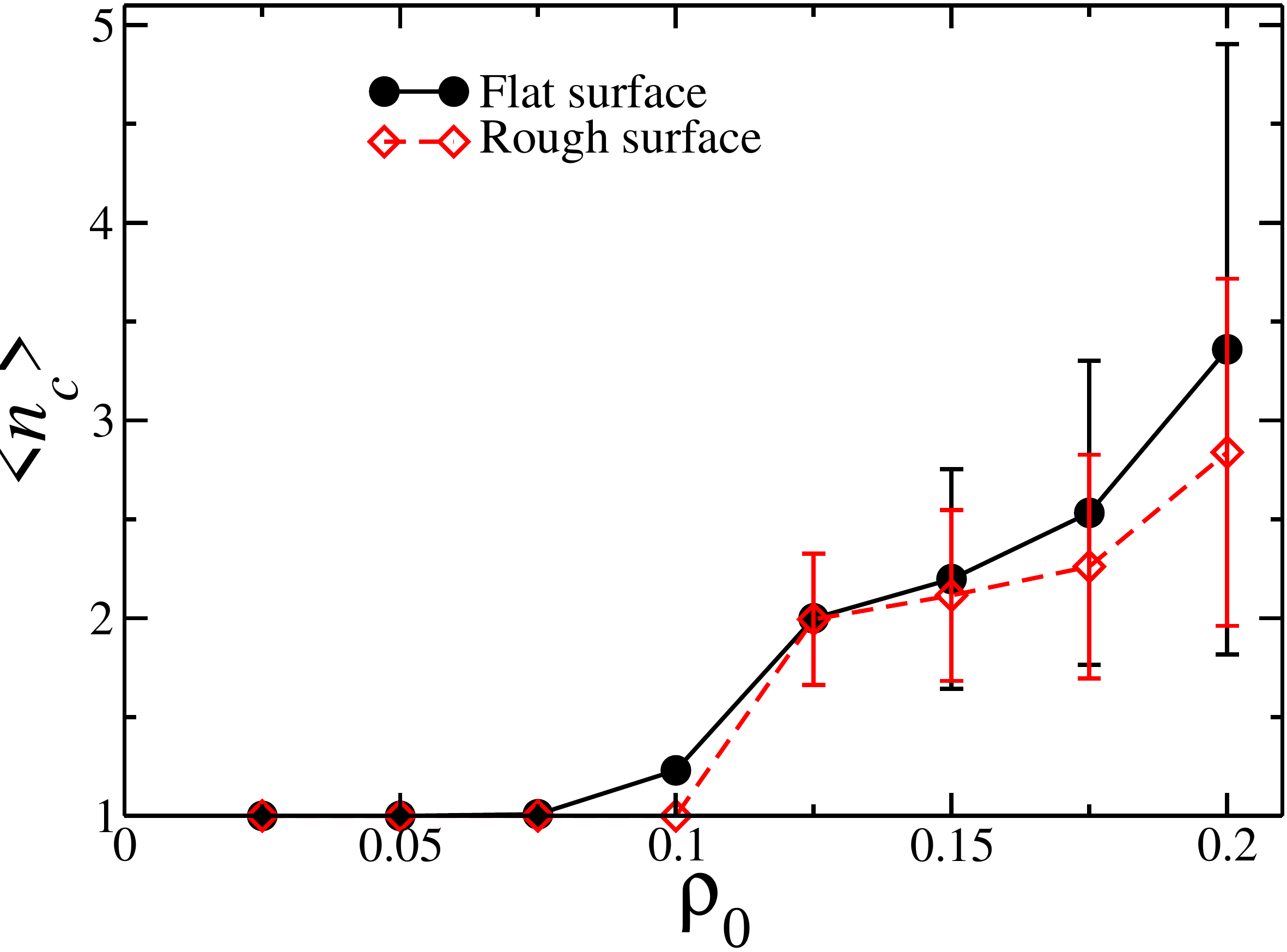}}
    \subfigure[]{\includegraphics[width=0.4\textwidth]{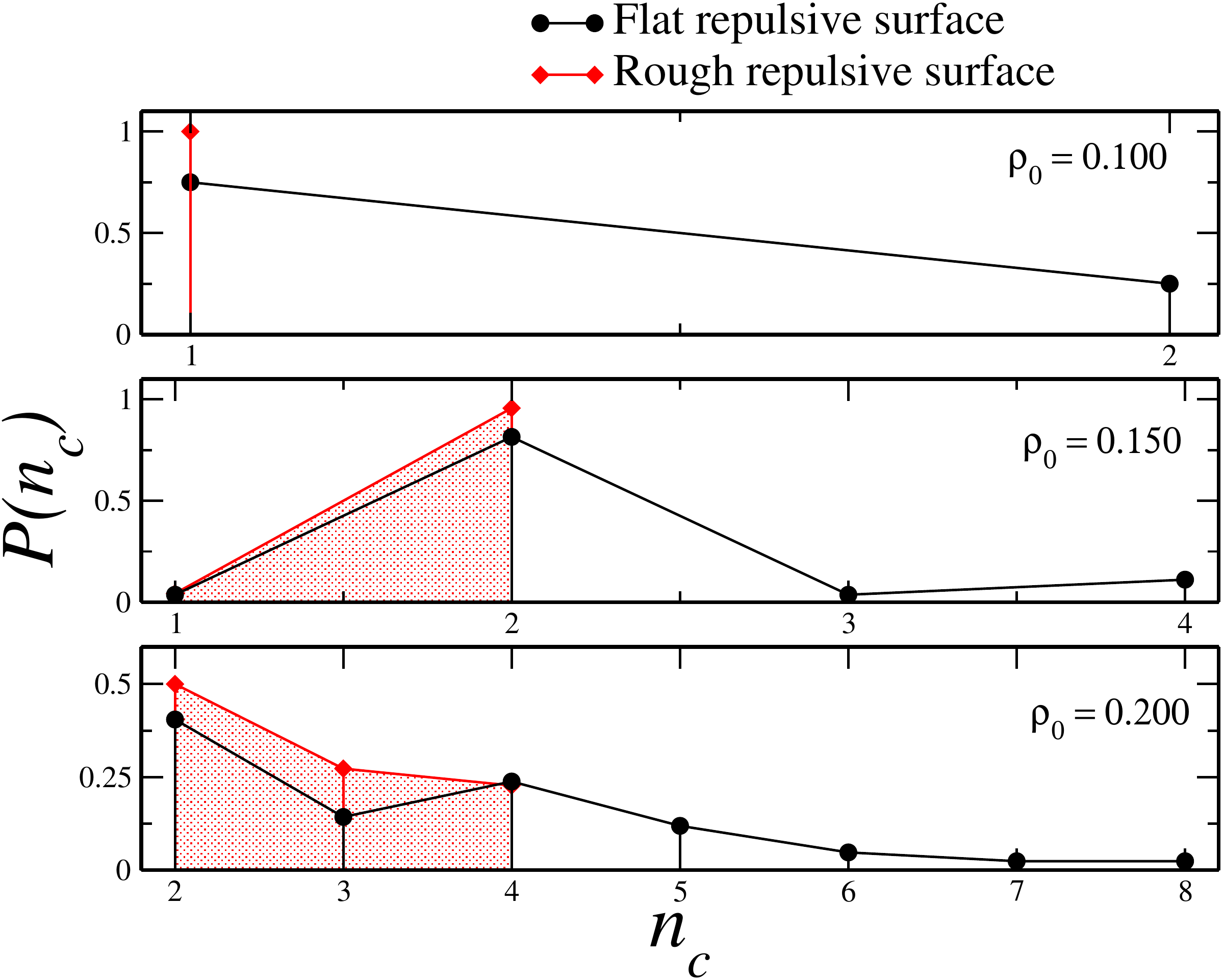}}
     \caption{(a) Ramp colloids adsorption isotherm for $T = 0.10$. (b) Layers (I), (II) and (III) structure at the density $\rho_0 = 0.100$ for flat and rough surfaces. LRDF of the contact layer (I) for all densities in (c) flat and (d) rough sufaces. (e) Mean number of particles in one cluster in the contact layer as function of bulk density and (f) probability of find a cluster with size $n_c$ at distinct values of bulk densities.}
        \label{fig3}
    \end{figure}

Is remarkable the layering observed at $T = 0.10$ and $\rho_0 = 0.100$. The fluid is organized in well defined planes along all the simulation box. This specific layering reflects in a anomalous adsorption behavior. In the figure~\ref{fig3}(a) we show the adsorption isotherm at $T = 0.10$ as function of the bulk density . As we can see, the density $\rho = 0.100$ corresponds to a maxima in the adsorption. From the 2D phase diagram obtained in our previous study~\cite{Marques20}, we know that this point is located in the solid hexagonal phase. Despite the fact that going from 2D to 3D or slab systems can change the phase diagram~\cite{Duda14,Fomin20,Bordin18,Krott13b}, each layer at this density is in a hexagonal lattice. Here is clear how the surface structure can affect the fluid layers structure. In the flat surface even the contact layer do not a perfect hexagonal structure, as we show in the upper panel of the figure~\ref{fig3}(b). As we move away from the wall more vacancies and defects can be seen. On the other hand, rough surfaces induces a well defined layering  without defects in the hexagonal lattice, as we can see in the bottom panel of the figure~\ref{fig3}(b).

We can see the defects by looking at the LRDF of the contact layer (I), shown in figure~\ref{fig3}(c) for flat walls and in figure~\ref{fig3}(d) for rough walls. As we can see, the overall behavior is similar: at low bulk densities the LRDF indicates no ordering, as in a gas-like phase. As $\rho_0$ increases, the peak in the LRDF near the second length scale, at $r_2 \approx 2.2$, grows and reach a maximum. At this point, the peak in the LRDF correspondent to the first length scale, at $r_1 \approx 1.2$, appears. However, for flat surfaces, we can see occupancy in the first length scale at $\rho_0 = 0.100$ while the rough surfaces has the higher occupancy in the second length scale and no particles at the first length scale. The occupancy in the first length scale indicates particles aggregation. This is corroborated by the analysis of the mean number of particle in one aggregate cluster at the contact layer (I), shown in the figure~\ref{fig3}(e). For flat surfaces and $\rho_0 = 0.100$ the mean cluster size is bigger than 1.0 - indicating the formation of aggregates, while for rough surfaces at $\rho_0 = 0.1$ no cluster were formed - the probability $P(n_c)$ shown in the upper panel of the figure~\ref{fig3}(f) also shows this. Increasing the density, the cluster size increases, and the occupancy in the first length scale increases as well.

    \setcounter{subfigure}{0}
    \begin{figure}[ht!]
        \centering
                \subfigure[]{\includegraphics[width=0.5\textwidth]{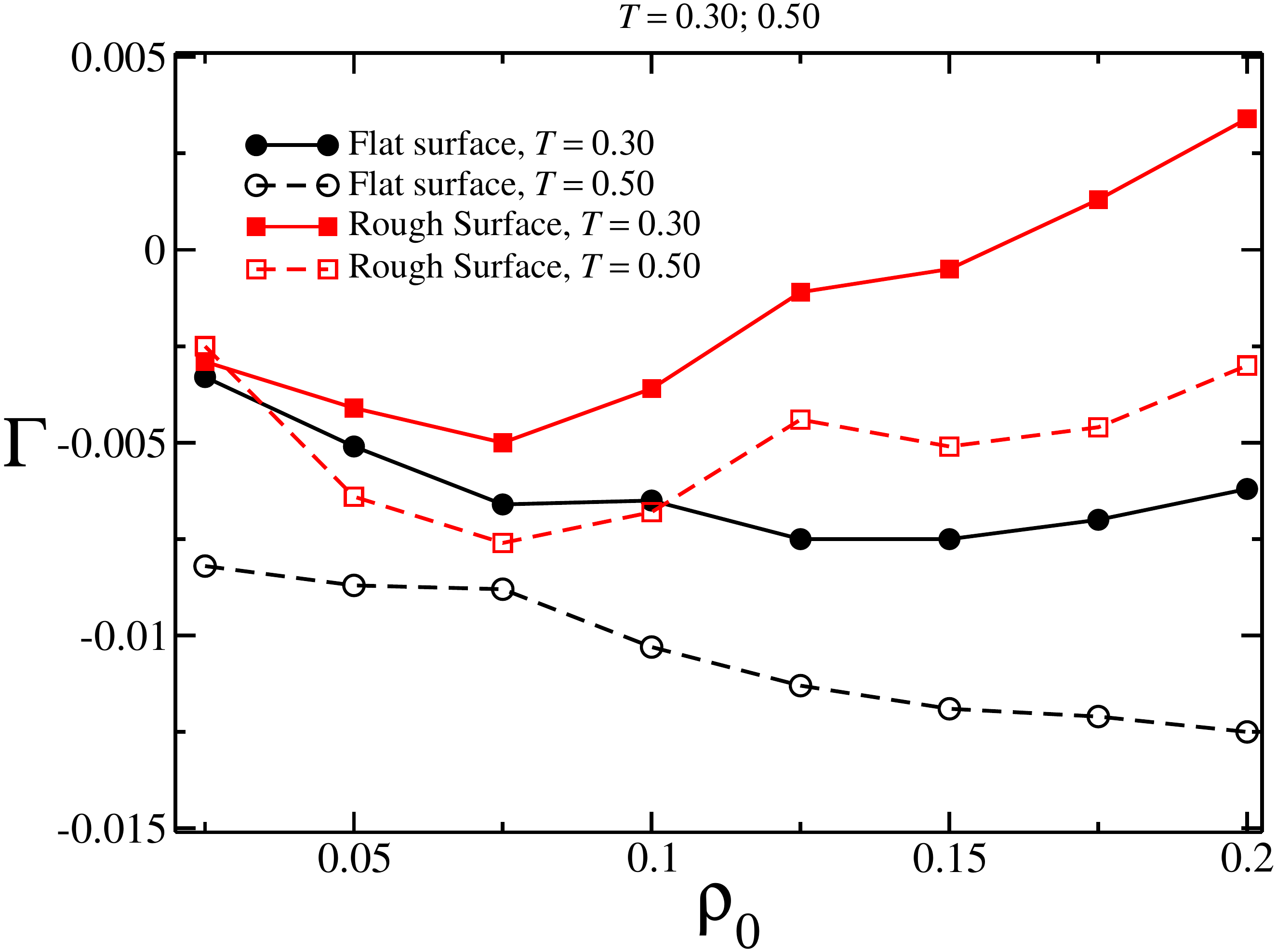}}
\subfigure[]{\includegraphics[width=0.45\textwidth]{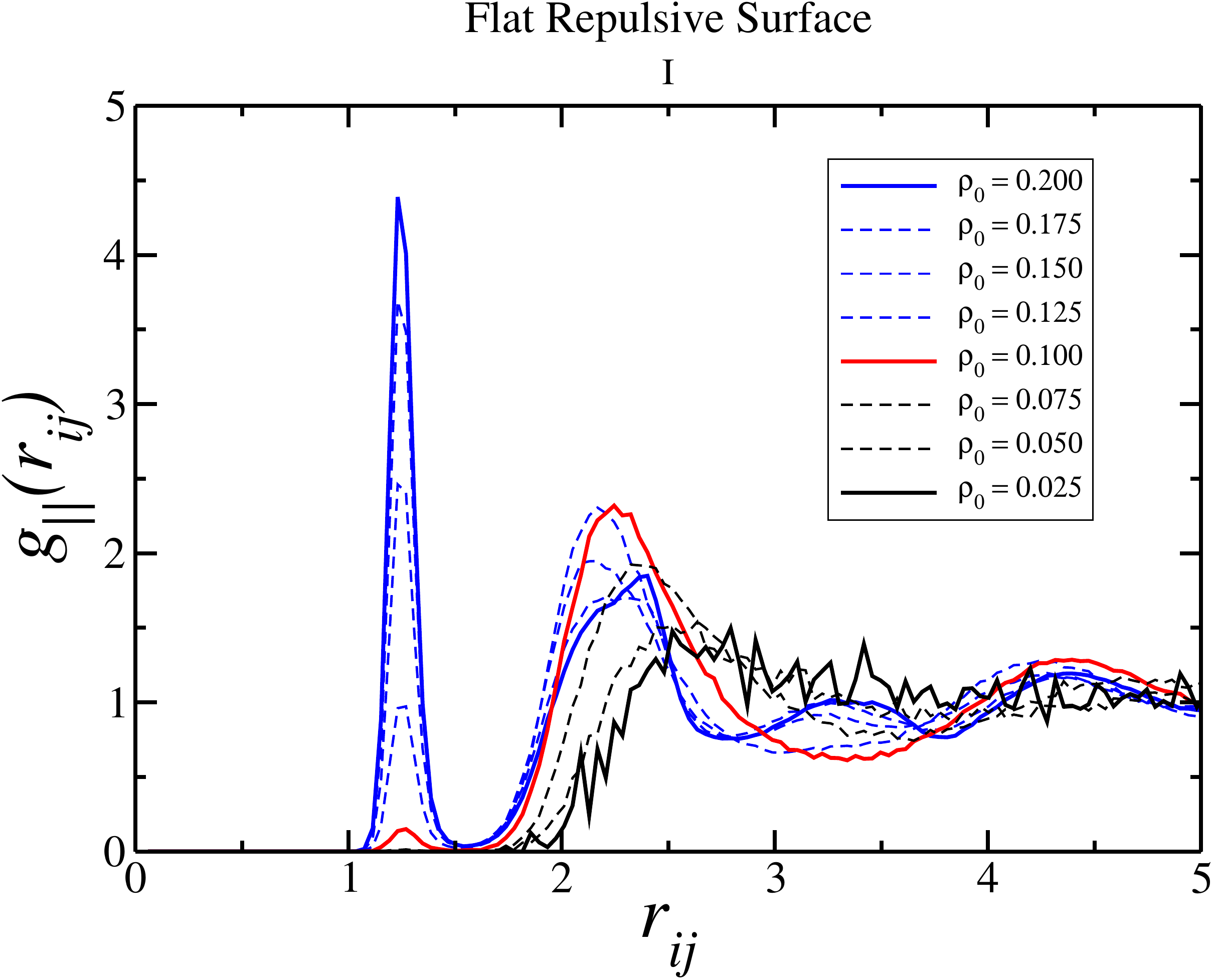}}
\subfigure[]{\includegraphics[width=0.45\textwidth]{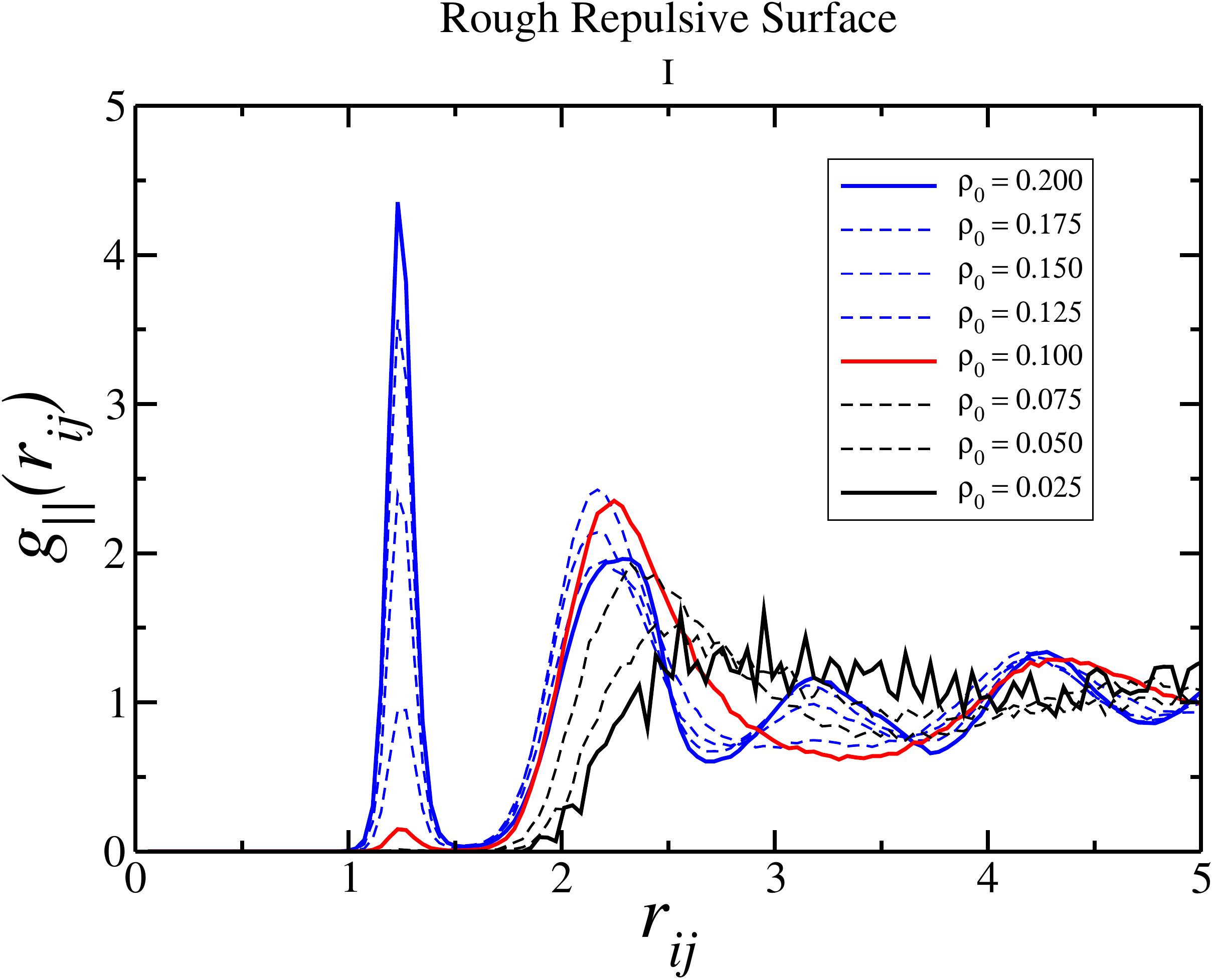}}
\subfigure[]{\includegraphics[width=0.425\textwidth]{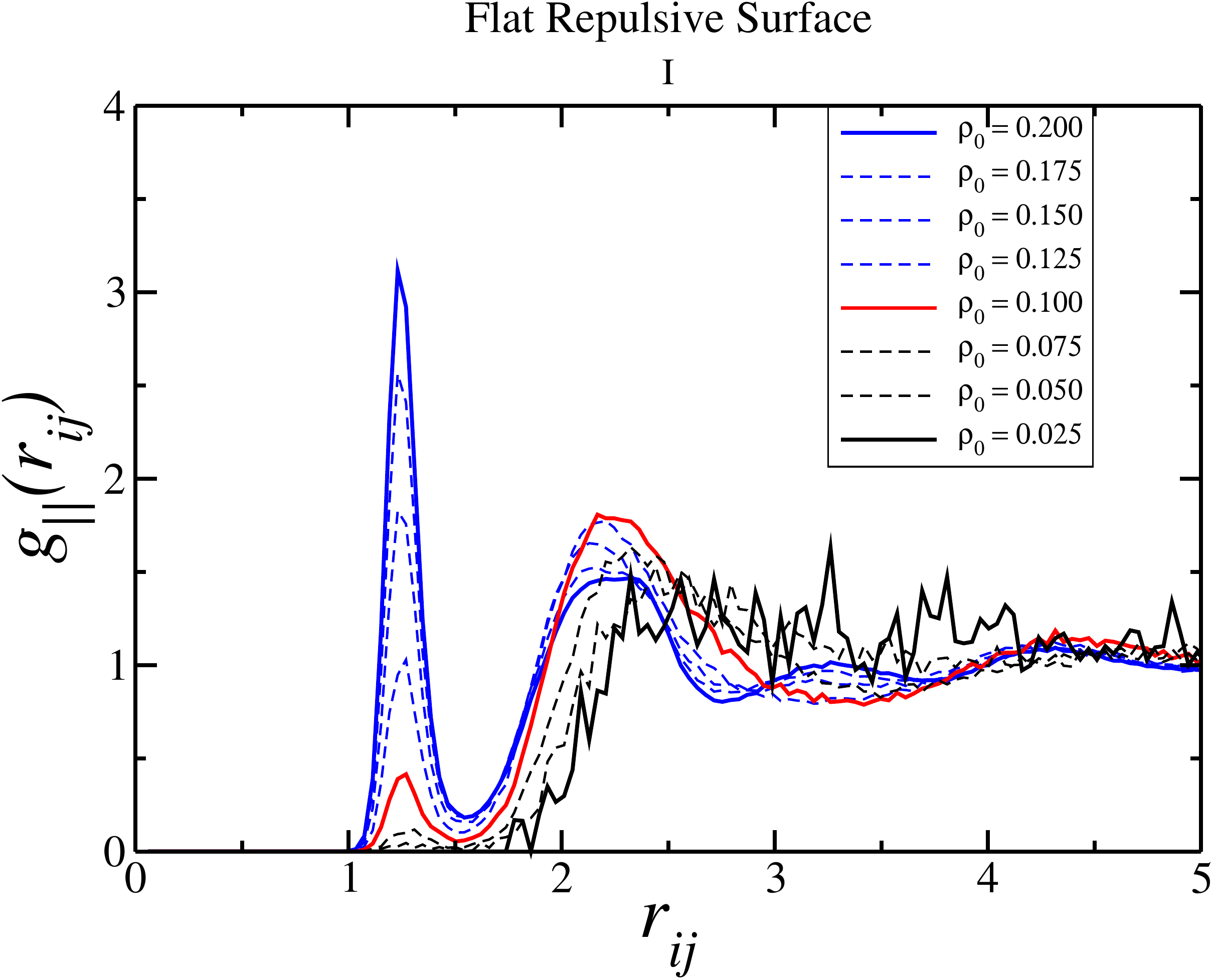}}
\subfigure[]{\includegraphics[width=0.425\textwidth]{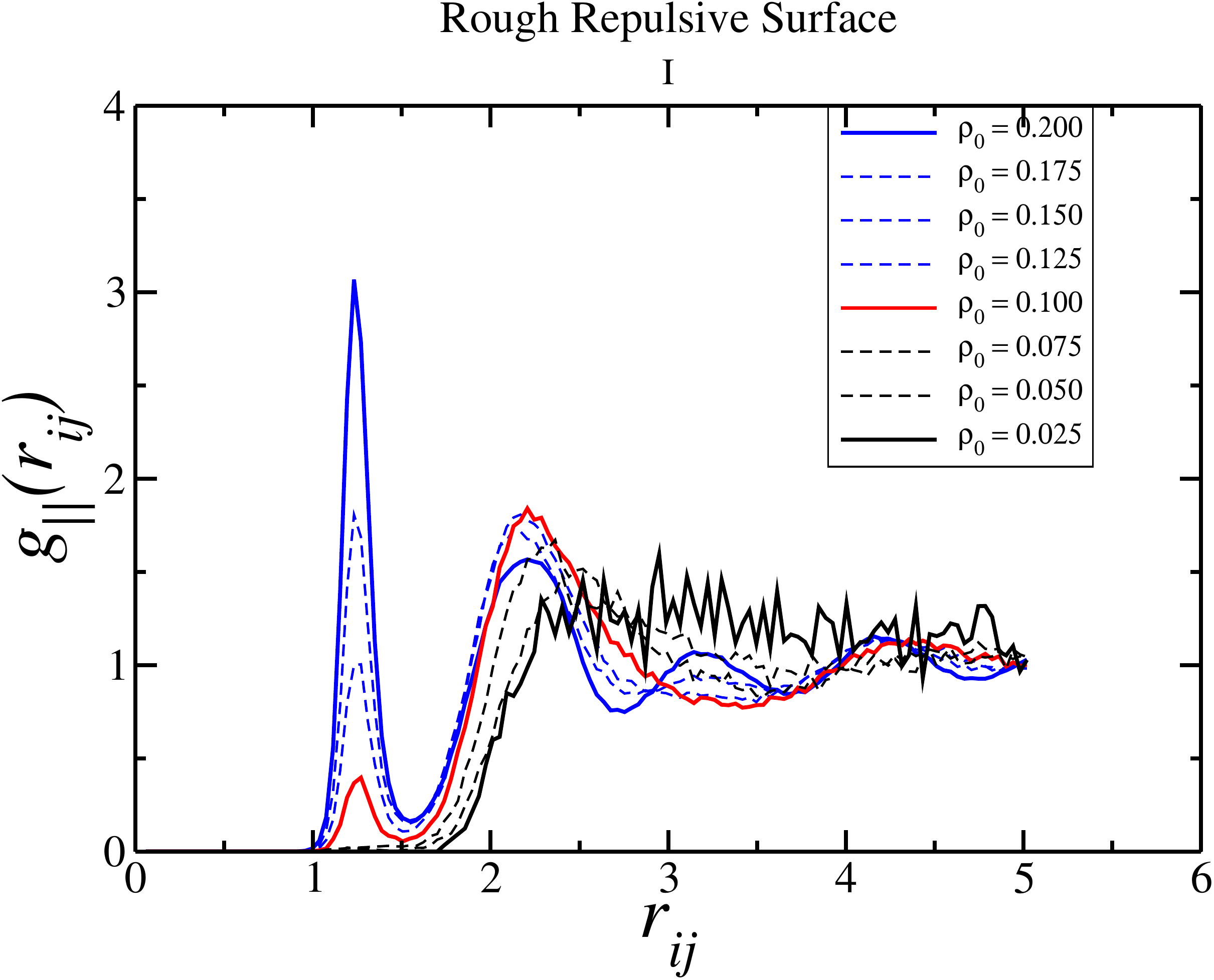}}

        \caption{Adsorption isotherms for $T = 0.30$ and $T = 0.50$ (a) and the LRDF for $T = 0.30$ and flat (b) or rough (c) adsorption surfaces and for $T = 0.50$ and flat (d) and or (e) adsorption surfaces}
        \label{fig4}
    \end{figure}

\begin{figure}[ht!]
        \centering
\subfigure[]{\includegraphics[width=0.575\textwidth]{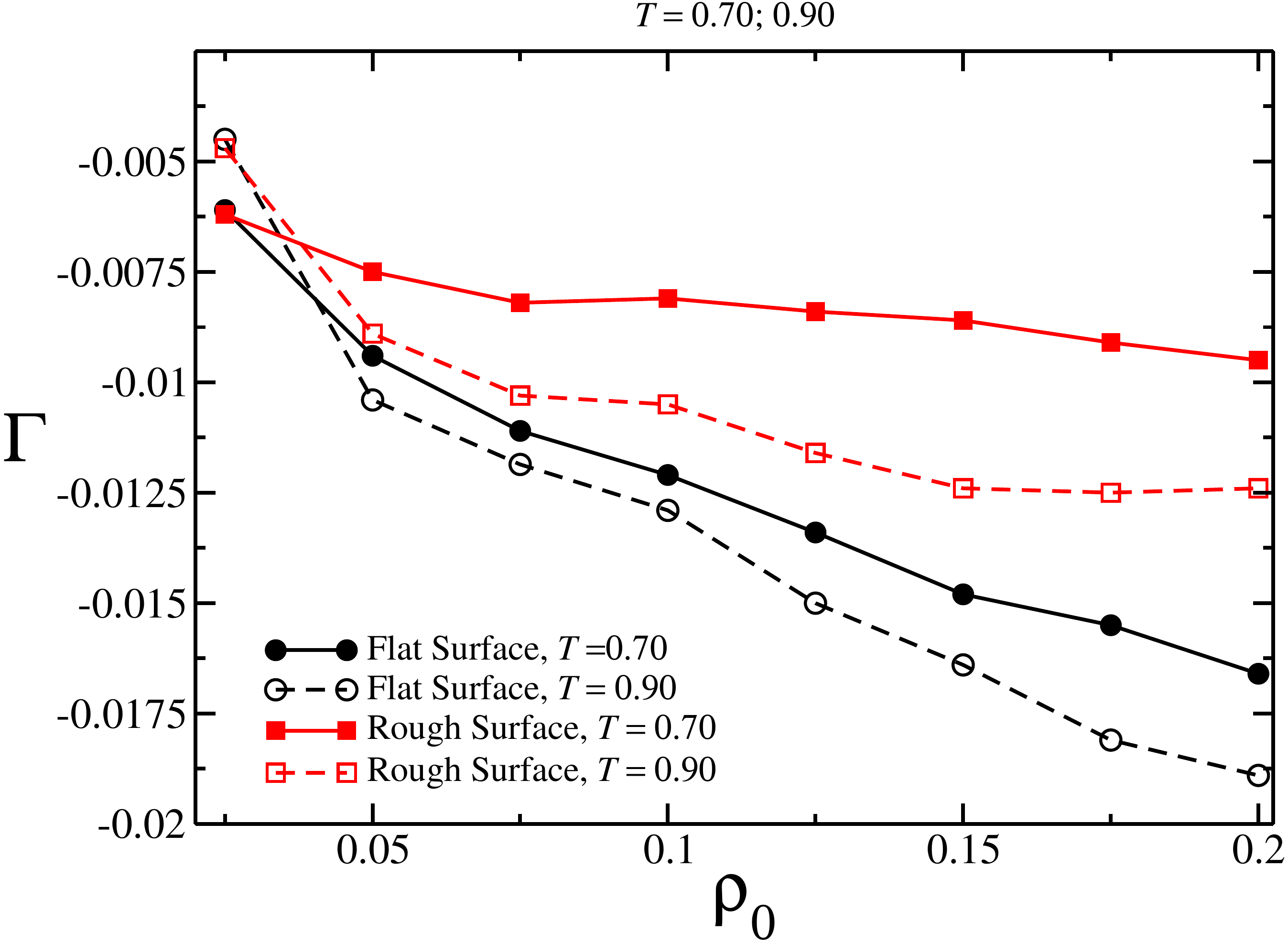}}
\subfigure[]{\includegraphics[width=0.425\textwidth]{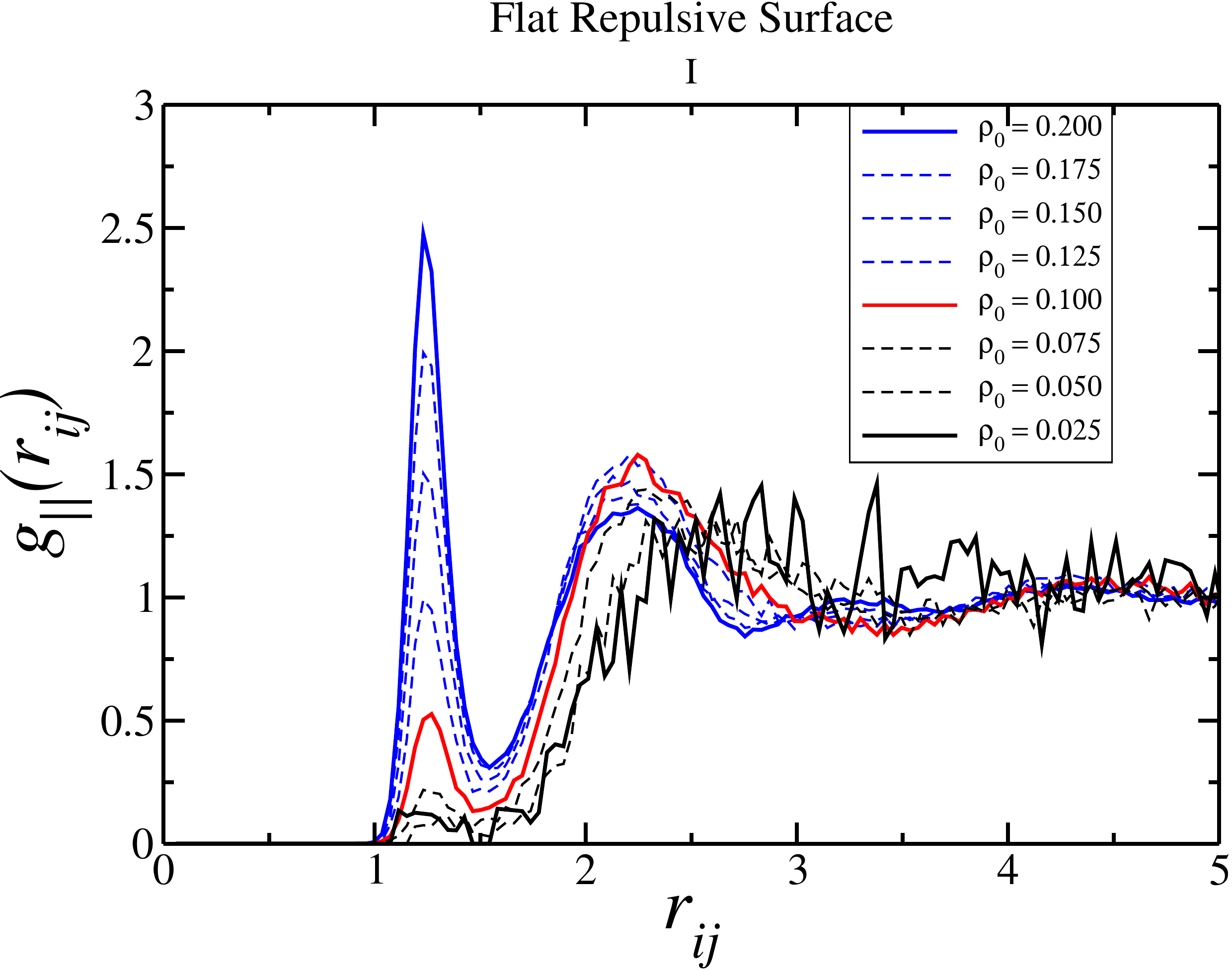}}
\subfigure[]{\includegraphics[width=0.425\textwidth]{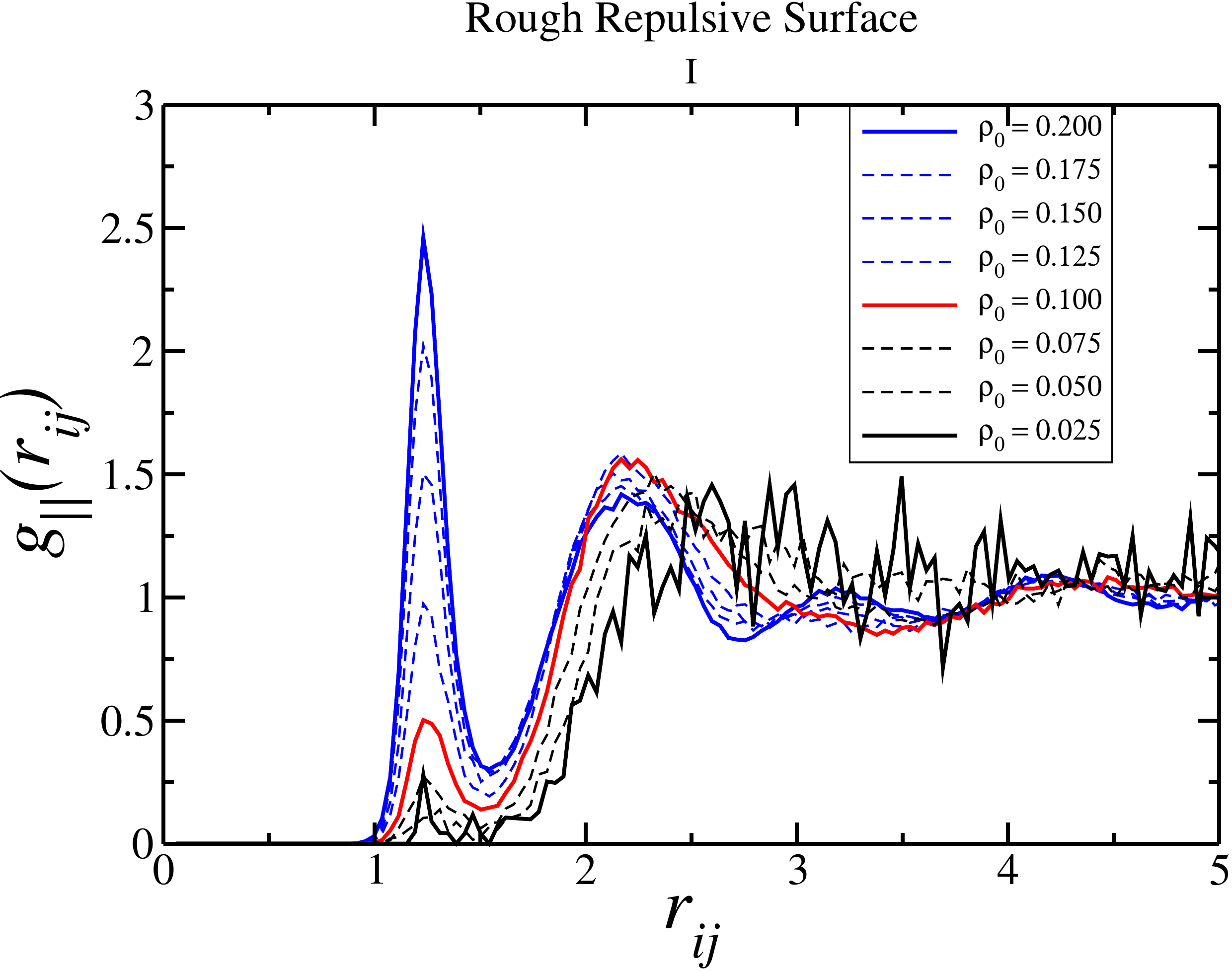}}
        \caption{Adsorption isotherms for $T = 0.70$ and $T = 0.00$ (a) and the LRDF for $T = 0.70$ and flat (c) or rough (d) adsorption surfaces.}
        \label{fig5}
    \end{figure}

Increasing $T$ in the 2D system~\cite{Marques20} melts the solid hexagonal phase. Here we observe the same. However, at the intermediate temperatures $T = 0.30$ and $T = 0.50$, the adsorption isotherms show an interesting behavior that can be related to the changes in the occupancy in each length scales. As we shown in figure~\ref{fig4}(a), for $T = 0.30$ both surfaces have the anomalous behavior in the adsorption curve, that can be related to the competition between the scales shown in figures~\ref{fig4}(b) and (c). For the isotherms at $T = 0.50$ the anomaly remains for the rough surface, but it practically vanishes for the flat surface, as we can see in the figure~\ref{fig4}(a). This distinct behavior is also related with the LRDF of the contact layer, shown in figures~\ref{fig4}(d) and (e). For the flat surfaces the first length scale is occupied even at low densities, since the thermal energy is high enough to overcome the ramp entalpic contribution for the total energy. On the other hand, for the rough adsorption surface there is the extra penalty due the friction with the surface beads. As consequence, the competition between the scales and the anomaly is observed for the rough case. So, heating up the system the entropic contribution for the free energy overcome the penalties from the ramp and from the friction, ending the competition, as we show in the figure~\ref{fig5}(b) and (c), and the adsorption anomaly, figure~\ref{fig5}(a), for $T = 0.90$. 

    \setcounter{subfigure}{0}

\begin{figure}[ht]
        \centering
    \includegraphics[width=0.55\textwidth]{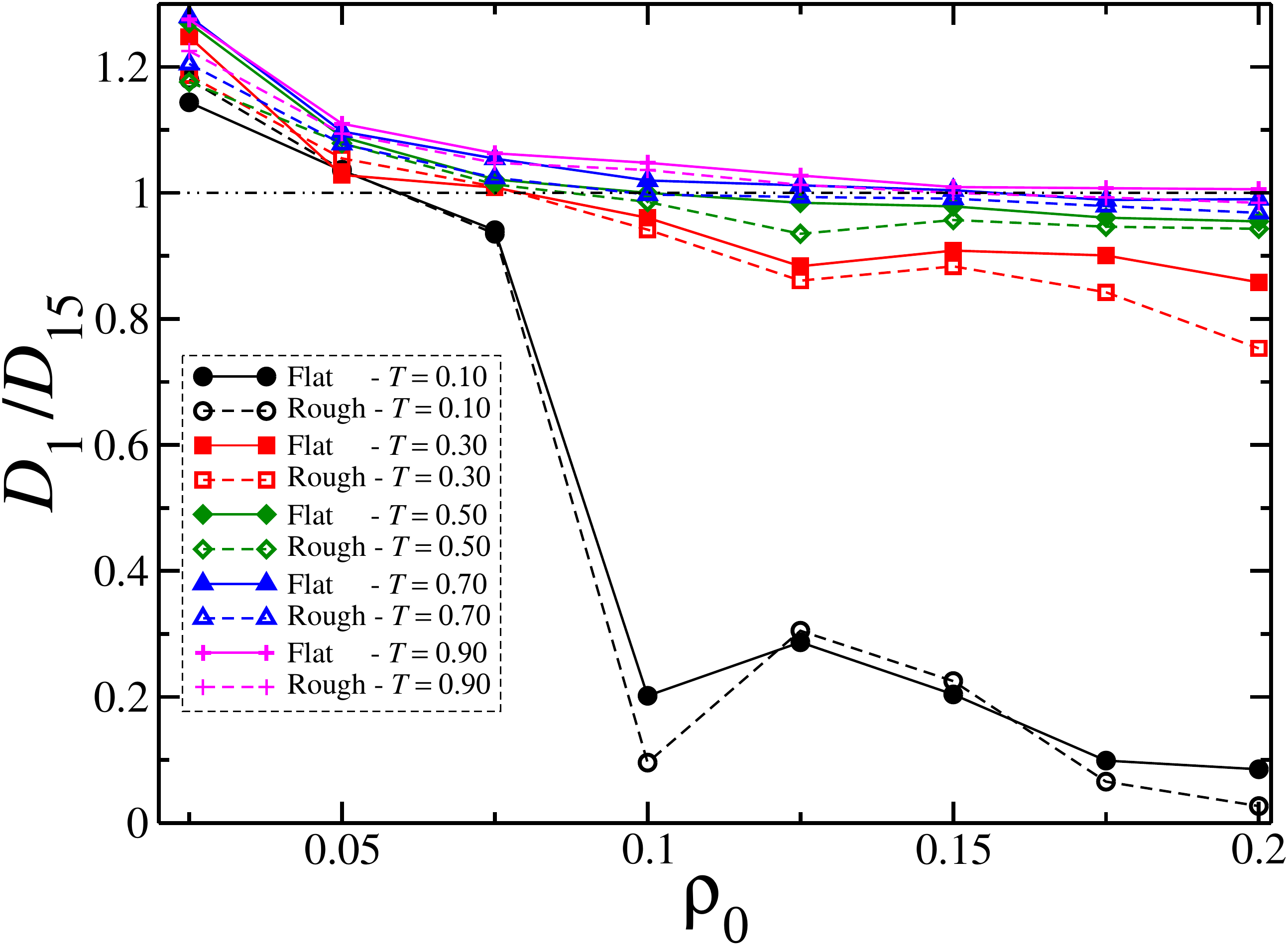}
        \caption{First layer diffusion divided by the central layer (bulk-like) diffusion for distinct bulk densities. }
        \label{fig6}
    \end{figure}

The competition between the scales observed in the LRDF can be related to the existence of water-like anomalies~\cite{Ney2009}. The diffusion anomaly is characterized by the increase in the self-diffusion constant $D$ as the pressure or density increases. When confined or near surfaces, this anomalous behavior is affected by the surface~\cite{Kohler19, Kohler18, Bordin17, Nogueira20}. To see how the colloids diffuses at the interface or in the bulk we have evaluated the lateral mean square displacement (LMSD) for distinct layer. With this, we obtain the lateral diffusion constant $D_l$ for each layer, with $l$ ranging from 1, the contact layer, to 15, the layer exactly at the simulation box center where the density was fixed in $\rho_0$. Here we show in the figure~\ref{fig6} the contact diffusion $D_1$ divided by the bulk diffusion $D_{15}$ for all the temperatures. Interesting, for all cases the contact layer diffuses faster at the interface when the colloidal density is small. This is counter-intuitive, once we should expect a smaller diffusion due the friction with the wall. However, core-softened fluids can show a anomalous increase in $D$ near solvophobic surfaces, as we have show previously~\cite{Bordin13, Bordin12b}. As $\rho_0$ increases, we can see that the isotherms have distinct behaviors. At high $T$, were no adsorption anomaly was observed, the diffusion decreases with the density. However, for the cases where the system has adsorption anomaly, we observe a diffusion anomaly. The ratio $D_1/D_{15}$ decay with $\rho_0$, indicating that the contact layer is diffuse slower than the bulk layers up to a threshold, where the curve has a mininum and increases as $\rho_0$ increases - similar to the water-like diffusion anomaly. This minima in the diffusion corresponds to the maxima observed for the adsorption, indicating that these quantities are related. This allow us to correlate the adsorption not only to the structure of the adsorbed colloids, but to their dynamic as well.

\section{Conclusions}
\label{Conclu}

In this paper we have explored the behavior of SALR particles near solid surfaces.  Two species of surfaces were simulated: a flat, smooth surface and a rough, structured one. By controlling the bulk density, we were able to see how the colloids adsorb at the surface at distinct temperatures. As usual for this system, we observe a layering as the density increases at low temperatures. Curiously, the system has more layers at the intermediary simulated density than at the higher density. This was related with the observation of a triangular lattice in the contact layer. This structural conformation was observed for all layers at $T = 0.10$ and $\rho_0 = 0.100$ - with defects and vacancies in the structure of the layers in the case of flat walls. 

This well defined layering at $T = 0.10$ and $\rho_0 = 0.100$ leads to a maxima in the NP adsorption. After that, the colloids start to aggregate. SALR interactions are characterized by the existence of two length scales, the short range attraction and the long range repulsion. This clustering is related to the occupancy in the first length scale - what was not observed prior to the adsorption maxima. Essentially, as $\rho_0$ increases the packing allow the colloids to overcome the energetic penalty from the long range repulsion. With this, they became closer and start to aggregate. Similarly, the temperatures $T = 0.30$ and $T = 0.50$ have changed in the adsorption curve when the particles starts to occupy the first characteristic distance.
However, when $T$ increases, the gain in kinetic energy allow the colloids to overcome the repulsive barrier even at low densities. With this, the first length scale is occupied even at low densities and no maxima is observed in the adsorption. 

To check how it is related to the dynamical properties, we evaluated the layer's diffusion and compare the diffusion of the contact layer and the central (bulk-like) layer. We could see that the maxima in the adsorption corresponds to a minima in the diffusion - this diffusion minima as function of density is similar to the waterlike diffusion anomaly, which is related to competition between the length scales observed in the water molecule interaction. 

With this, we could connect adsorption, clustering and diffusion with the competition between the length scales of SALR. In fact, the correlation between maximum in the adsorption and clustering agrees with the recent work by Bildanau and co-workers for 2D systems~\cite{ciach2020}, and our results provide information regarding the connection with the diffusion behavior at the interface and the competition between the scales.

\begin{acknowledgement}

Without public funding this research would be impossible. MSM thanks the Brazilian Agencies Conselho Nacional de Desenvolvimento Cient\'ifico e Tecnol\'ogico (CNPq) for the PhD Scholarship and Coordena\c c\~ao de Aperfei\c coamento de Pessoal de N\'ivel Superior (CAPES) for the support to the collaborative period in the Instituto de Química Fisica Rocasolano. JRB acknowledge the Brazilian agencies CNPq and Funda\c c\~ao de Apoio a Pesquisa do Rio Grande do Sul (FAPERGS) for financial support. JRB is greatly indebted to Alexandre Diehl for illuminating discussions. All simulations were performed in the SATOLEP Cluster of the Group of Theory and Simulation in Complex Systems from UFPel. 

\end{acknowledgement}


\begin{mcitethebibliography}{71}
\providecommand*\natexlab[1]{#1}
\providecommand*\mciteSetBstSublistMode[1]{}
\providecommand*\mciteSetBstMaxWidthForm[2]{}
\providecommand*\mciteBstWouldAddEndPuncttrue
  {\def\EndOfBibitem{\unskip.}}
\providecommand*\mciteBstWouldAddEndPunctfalse
  {\let\EndOfBibitem\relax}
\providecommand*\mciteSetBstMidEndSepPunct[3]{}
\providecommand*\mciteSetBstSublistLabelBeginEnd[3]{}
\providecommand*\EndOfBibitem{}
\mciteSetBstSublistMode{f}
\mciteSetBstMaxWidthForm{subitem}{(\alph{mcitesubitemcount})}
\mciteSetBstSublistLabelBeginEnd
  {\mcitemaxwidthsubitemform\space}
  {\relax}
  {\relax}

\bibitem[Dąbrowski(2001)]{dabrowski2001}
Dąbrowski,~A. Adsorption — from theory to practice. \emph{Advances in
  Colloid and Interface Science} \textbf{2001}, \emph{93}, 135--224\relax
\mciteBstWouldAddEndPuncttrue
\mciteSetBstMidEndSepPunct{\mcitedefaultmidpunct}
{\mcitedefaultendpunct}{\mcitedefaultseppunct}\relax
\EndOfBibitem
\bibitem[col(2016)]{colloid_chapter}
\emph{Introduction to Applied Colloid and Surface Chemistry}; John Wiley and
  Sons, Ltd, 2016; Chapter 7, pp 161--184\relax
\mciteBstWouldAddEndPuncttrue
\mciteSetBstMidEndSepPunct{\mcitedefaultmidpunct}
{\mcitedefaultendpunct}{\mcitedefaultseppunct}\relax
\EndOfBibitem
\bibitem[Bayliss(1911)]{bayliss1911}
Bayliss,~W.~M. The Properties of Colloidal Systems.-II. On Adsorption as
  Preliminary to Chemical Reaction. \emph{Proceedings of the Royal Society of
  London. Series B, Containing Papers of a Biological Character} \textbf{1911},
  \emph{84}, 81--98\relax
\mciteBstWouldAddEndPuncttrue
\mciteSetBstMidEndSepPunct{\mcitedefaultmidpunct}
{\mcitedefaultendpunct}{\mcitedefaultseppunct}\relax
\EndOfBibitem
\bibitem[Vold and Sivaramakrishnan(1958)Vold, and Sivaramakrishnan]{vold1958}
Vold,~R.~D.; Sivaramakrishnan,~N.~H. The Origin of the Maximum in the
  Adsorption Isotherms of Association Colloids. \emph{The Journal of Physical
  Chemistry} \textbf{1958}, \emph{62}, 984--989\relax
\mciteBstWouldAddEndPuncttrue
\mciteSetBstMidEndSepPunct{\mcitedefaultmidpunct}
{\mcitedefaultendpunct}{\mcitedefaultseppunct}\relax
\EndOfBibitem
\bibitem[Robens and Jayaweera(2014)Robens, and Jayaweera]{robens2014}
Robens,~E.; Jayaweera,~S. A.~A. Early History of Adsorption Measurements.
  \emph{Adsorption Science \& Technology} \textbf{2014}, \emph{32},
  425--442\relax
\mciteBstWouldAddEndPuncttrue
\mciteSetBstMidEndSepPunct{\mcitedefaultmidpunct}
{\mcitedefaultendpunct}{\mcitedefaultseppunct}\relax
\EndOfBibitem
\bibitem[Tareq \latin{et~al.}(2019)Tareq, Akter, and Azam]{tareq2019}
Tareq,~R.; Akter,~N.; Azam,~M.~S. In \emph{Biochar from Biomass and Waste};
  Ok,~Y.~S., Tsang,~D.~C., Bolan,~N., Novak,~J., Eds.; Elsevier, 2019; pp 169
  -- 209\relax
\mciteBstWouldAddEndPuncttrue
\mciteSetBstMidEndSepPunct{\mcitedefaultmidpunct}
{\mcitedefaultendpunct}{\mcitedefaultseppunct}\relax
\EndOfBibitem
\bibitem[Tang(2019)]{tang2019}
Tang,~F. \emph{Structures and Dynamics of Interfacial Water: Input from
  Theoretical Vibrational Sum-frequency Spectroscopy}; Springer Theses;
  Springer Singapore, 2019\relax
\mciteBstWouldAddEndPuncttrue
\mciteSetBstMidEndSepPunct{\mcitedefaultmidpunct}
{\mcitedefaultendpunct}{\mcitedefaultseppunct}\relax
\EndOfBibitem
\bibitem[Ciach \latin{et~al.}(2013)Ciach, Pękalski, and Góźdź]{ciach2013}
Ciach,~A.; Pękalski,~J.; Góźdź,~W.~T. Origin of similarity of phase
  diagrams in amphiphilic and colloidal systems with competing interactions.
  \emph{Soft Matter} \textbf{2013}, \emph{9}, 6301--6308\relax
\mciteBstWouldAddEndPuncttrue
\mciteSetBstMidEndSepPunct{\mcitedefaultmidpunct}
{\mcitedefaultendpunct}{\mcitedefaultseppunct}\relax
\EndOfBibitem
\bibitem[Campbell \latin{et~al.}(2005)Campbell, Anderson, van Duijneveldt, and
  Bartlett]{bartlett2005}
Campbell,~A.~I.; Anderson,~V.~J.; van Duijneveldt,~J.~S.; Bartlett,~P.
  Dynamical Arrest in Attractive Colloids: The Effect of Long-Range Repulsion.
  \emph{Phys. Rev. Lett.} \textbf{2005}, \emph{94}, 208301\relax
\mciteBstWouldAddEndPuncttrue
\mciteSetBstMidEndSepPunct{\mcitedefaultmidpunct}
{\mcitedefaultendpunct}{\mcitedefaultseppunct}\relax
\EndOfBibitem
\bibitem[Adamczyk \latin{et~al.}(2002)Adamczyk, Weroński, and
  Musiał]{adam2002}
Adamczyk,~Z.; Weroński,~P.; Musiał,~E. Colloid Particle Adsorption at Random
  Site (Heterogeneous) Surfaces. \emph{Journal of Colloid and Interface
  Science} \textbf{2002}, \emph{248}, 67 -- 75\relax
\mciteBstWouldAddEndPuncttrue
\mciteSetBstMidEndSepPunct{\mcitedefaultmidpunct}
{\mcitedefaultendpunct}{\mcitedefaultseppunct}\relax
\EndOfBibitem
\bibitem[Adamczyk(2012)]{adam2012}
Adamczyk,~Z. Modeling adsorption of colloids and proteins. \emph{Current
  Opinion in Colloid \& Interface Science} \textbf{2012}, \emph{17}, 173 --
  186\relax
\mciteBstWouldAddEndPuncttrue
\mciteSetBstMidEndSepPunct{\mcitedefaultmidpunct}
{\mcitedefaultendpunct}{\mcitedefaultseppunct}\relax
\EndOfBibitem
\bibitem[Wasilewska and Adamczyk(2011)Wasilewska, and Adamczyk]{monika2011}
Wasilewska,~M.; Adamczyk,~Z. Fibrinogen Adsorption on Mica Studied by AFM and
  in Situ Streaming Potential Measurements. \emph{Langmuir} \textbf{2011},
  \emph{27}, 686--696, PMID: 21155546\relax
\mciteBstWouldAddEndPuncttrue
\mciteSetBstMidEndSepPunct{\mcitedefaultmidpunct}
{\mcitedefaultendpunct}{\mcitedefaultseppunct}\relax
\EndOfBibitem
\bibitem[Rahmani \latin{et~al.}(2016)Rahmani, Wang, Manoharan, and
  Colosqui]{amir2016}
Rahmani,~A.~M.; Wang,~A.; Manoharan,~V.~N.; Colosqui,~C.~E. Colloidal particle
  adsorption at liquid interfaces: capillary driven dynamics and thermally
  activated kinetics. \emph{Soft Matter} \textbf{2016}, \emph{12},
  6365--6372\relax
\mciteBstWouldAddEndPuncttrue
\mciteSetBstMidEndSepPunct{\mcitedefaultmidpunct}
{\mcitedefaultendpunct}{\mcitedefaultseppunct}\relax
\EndOfBibitem
\bibitem[Imperio and Reatto(2007)Imperio, and Reatto]{imperio2007}
Imperio,~A.; Reatto,~L. Microphase morphology in two-dimensional fluids under
  lateral confinement. \emph{Phys. Rev. E} \textbf{2007}, \emph{76},
  040402\relax
\mciteBstWouldAddEndPuncttrue
\mciteSetBstMidEndSepPunct{\mcitedefaultmidpunct}
{\mcitedefaultendpunct}{\mcitedefaultseppunct}\relax
\EndOfBibitem
\bibitem[Rabe \latin{et~al.}(2011)Rabe, Verdes, and Seeger]{rabe2011}
Rabe,~M.; Verdes,~D.; Seeger,~S. Understanding protein adsorption phenomena at
  solid surfaces. \emph{Advances in Colloid and Interface Science}
  \textbf{2011}, \emph{162}, 87 -- 106\relax
\mciteBstWouldAddEndPuncttrue
\mciteSetBstMidEndSepPunct{\mcitedefaultmidpunct}
{\mcitedefaultendpunct}{\mcitedefaultseppunct}\relax
\EndOfBibitem
\bibitem[Stradner and Schurtenberger(2020)Stradner, and
  Schurtenberger]{stradner2020}
Stradner,~A.; Schurtenberger,~P. Potential and limits of a colloid approach to
  protein solutions. \emph{Soft Matter} \textbf{2020}, \emph{16},
  307--323\relax
\mciteBstWouldAddEndPuncttrue
\mciteSetBstMidEndSepPunct{\mcitedefaultmidpunct}
{\mcitedefaultendpunct}{\mcitedefaultseppunct}\relax
\EndOfBibitem
\bibitem[Santos \latin{et~al.}(2017)Santos, Pekalski, and
  Panagiotopoulos]{santos2017}
Santos,~A.~P.; Pekalski,~J.; Panagiotopoulos,~A.~Z. Thermodynamic signatures
  and cluster properties of self-assembly in systems with competing
  interactions. \emph{Soft Matter} \textbf{2017}, \emph{13}, 8055--8063\relax
\mciteBstWouldAddEndPuncttrue
\mciteSetBstMidEndSepPunct{\mcitedefaultmidpunct}
{\mcitedefaultendpunct}{\mcitedefaultseppunct}\relax
\EndOfBibitem
\bibitem[Verwey \latin{et~al.}(1948)Verwey, Overbeek, and van Nes]{verwey1948}
Verwey,~E.; Overbeek,~J.; van Nes,~K. \emph{Theory of the Stability of
  Lyophobic Colloids: The Interaction of Sol Particles Having an Electric
  Double Layer}; Elsevier Publishing Company, 1948\relax
\mciteBstWouldAddEndPuncttrue
\mciteSetBstMidEndSepPunct{\mcitedefaultmidpunct}
{\mcitedefaultendpunct}{\mcitedefaultseppunct}\relax
\EndOfBibitem
\bibitem[Asakura and Oosawa(1954)Asakura, and Oosawa]{asakura1954}
Asakura,~S.; Oosawa,~F. On Interaction between Two Bodies Immersed in a
  Solution of Macromolecules. \emph{The Journal of Chemical Physics}
  \textbf{1954}, \emph{22}, 1255--1256\relax
\mciteBstWouldAddEndPuncttrue
\mciteSetBstMidEndSepPunct{\mcitedefaultmidpunct}
{\mcitedefaultendpunct}{\mcitedefaultseppunct}\relax
\EndOfBibitem
\bibitem[Asakura and Oosawa(1958)Asakura, and Oosawa]{asakura1958}
Asakura,~S.; Oosawa,~F. Interaction between particles suspended in solutions of
  macromolecules. \emph{Journal of Polymer Science} \textbf{1958}, \emph{33},
  183--192\relax
\mciteBstWouldAddEndPuncttrue
\mciteSetBstMidEndSepPunct{\mcitedefaultmidpunct}
{\mcitedefaultendpunct}{\mcitedefaultseppunct}\relax
\EndOfBibitem
\bibitem[Royall(2018)]{royall2018}
Royall,~C.~P. Hunting mermaids in real space: known knowns{,} known unknowns
  and unknown unknowns. \emph{Soft Matter} \textbf{2018}, \emph{14},
  4020--4028\relax
\mciteBstWouldAddEndPuncttrue
\mciteSetBstMidEndSepPunct{\mcitedefaultmidpunct}
{\mcitedefaultendpunct}{\mcitedefaultseppunct}\relax
\EndOfBibitem
\bibitem[Shukla \latin{et~al.}(2008)Shukla, Mylonas, Di~Cola, Finet, Timmins,
  Narayanan, and Svergun]{Shukla08}
Shukla,~A.; Mylonas,~E.; Di~Cola,~E.; Finet,~S.; Timmins,~P.; Narayanan,~T.;
  Svergun,~D.~I. Absence of equilibrium cluster phase in concentrated lysozyme
  solutions. \emph{Proceedings of the National Academy of Sciences}
  \textbf{2008}, \emph{105}, 5075--5080\relax
\mciteBstWouldAddEndPuncttrue
\mciteSetBstMidEndSepPunct{\mcitedefaultmidpunct}
{\mcitedefaultendpunct}{\mcitedefaultseppunct}\relax
\EndOfBibitem
\bibitem[Somerville \latin{et~al.}(2020)Somerville, Law, Rey, Vogel, Archer,
  and Buzza]{Somerville20}
Somerville,~W. R.~C.; Law,~A.~D.; Rey,~M.; Vogel,~N.; Archer,~A.~J.; Buzza,~D.
  M.~A. Pattern formation in two-dimensional hard-core/soft-shell systems with
  variable soft shell profiles. \emph{Soft Matter} \textbf{2020}, \emph{16},
  3564--3573\relax
\mciteBstWouldAddEndPuncttrue
\mciteSetBstMidEndSepPunct{\mcitedefaultmidpunct}
{\mcitedefaultendpunct}{\mcitedefaultseppunct}\relax
\EndOfBibitem
\bibitem[Cardoso \latin{et~al.}(2021)Cardoso, Hernandes, Nogueira, and
  Bordin]{Cardoso21}
Cardoso,~D.~S.; Hernandes,~V.~F.; Nogueira,~T.; Bordin,~J.~R. Structural
  behavior of a two length scale core-softened fluid in two dimensions.
  \emph{Physica A: Statistical Mechanics and its Applications} \textbf{2021},
  \emph{566}, 125628\relax
\mciteBstWouldAddEndPuncttrue
\mciteSetBstMidEndSepPunct{\mcitedefaultmidpunct}
{\mcitedefaultendpunct}{\mcitedefaultseppunct}\relax
\EndOfBibitem
\bibitem[Pekalski \latin{et~al.}(2014)Pekalski, Ciach, and
  Almarza]{almarza2014}
Pekalski,~J.; Ciach,~A.; Almarza,~N.~G. Periodic ordering of clusters and
  stripes in a two-dimensional lattice model. I. Ground state, mean-field phase
  diagram and structure of the disordered phases. \emph{The Journal of Chemical
  Physics} \textbf{2014}, \emph{140}, 114701\relax
\mciteBstWouldAddEndPuncttrue
\mciteSetBstMidEndSepPunct{\mcitedefaultmidpunct}
{\mcitedefaultendpunct}{\mcitedefaultseppunct}\relax
\EndOfBibitem
\bibitem[Ong \latin{et~al.}(2015)Ong, Williams, Singh, Schaible, Helms, and
  Milliron]{Ong15}
Ong,~G.~K.; Williams,~T.~E.; Singh,~A.; Schaible,~E.; Helms,~B.~A.;
  Milliron,~D.~J. Ordering in Polymer Micelle-Directed Assemblies of Colloidal
  Nanocrystals. \emph{Nano Letters} \textbf{2015}, \emph{15}, 8240--8244\relax
\mciteBstWouldAddEndPuncttrue
\mciteSetBstMidEndSepPunct{\mcitedefaultmidpunct}
{\mcitedefaultendpunct}{\mcitedefaultseppunct}\relax
\EndOfBibitem
\bibitem[Montes-Campos \latin{et~al.}(2017)Montes-Campos, Otero-Mato,
  Mendez-Morales, Cabeza, Gallego, Ciach, and Varela]{Campos17}
Montes-Campos,~H.; Otero-Mato,~J.~M.; Mendez-Morales,~T.; Cabeza,~O.;
  Gallego,~L.~J.; Ciach,~A.; Varela,~L.~M. Two-dimensional pattern formation in
  ionic liquids confined between graphene walls. \emph{Phys. Chem. Chem. Phys.}
  \textbf{2017}, \emph{19}, --\relax
\mciteBstWouldAddEndPuncttrue
\mciteSetBstMidEndSepPunct{\mcitedefaultmidpunct}
{\mcitedefaultendpunct}{\mcitedefaultseppunct}\relax
\EndOfBibitem
\bibitem[Quesada-Perez \latin{et~al.}(2001)Quesada-Perez, Moncho-Jorda,
  Martinez-Lopez, and Hidalgo-\'Alvarez]{colloid1}
Quesada-Perez,~M.; Moncho-Jorda,~A.; Martinez-Lopez,~F.; Hidalgo-\'Alvarez,~R.
  Probing interaction forces in colloidal monolayers: Inversion of structural
  data. \emph{Journal of Chemical Physics} \textbf{2001}, \emph{115},
  10897\relax
\mciteBstWouldAddEndPuncttrue
\mciteSetBstMidEndSepPunct{\mcitedefaultmidpunct}
{\mcitedefaultendpunct}{\mcitedefaultseppunct}\relax
\EndOfBibitem
\bibitem[{Contreras-Aburto} and amd R.C.~Priego(2010){Contreras-Aburto}, and
  amd R.C.~Priego]{colloid2}
{Contreras-Aburto},~C.; amd R.C.~Priego,~J.~M. Structure and effective
  interactions in parallel monolayers of charged spherical colloids.
  \emph{Journal of Chemical Physics} \textbf{2010}, \emph{132}, 174111\relax
\mciteBstWouldAddEndPuncttrue
\mciteSetBstMidEndSepPunct{\mcitedefaultmidpunct}
{\mcitedefaultendpunct}{\mcitedefaultseppunct}\relax
\EndOfBibitem
\bibitem[Haddadi \latin{et~al.}(2020)Haddadi, Skepö, Jannasch, Manner, and
  Forsman]{Haddadi20}
Haddadi,~S.; Skepö,~M.; Jannasch,~P.; Manner,~S.; Forsman,~J. Building
  polymer-like clusters from colloidal particles with isotropic interactions,
  in aqueous solution. \emph{Journal of Colloid and Interface Science}
  \textbf{2020}, \emph{581}, 669--681\relax
\mciteBstWouldAddEndPuncttrue
\mciteSetBstMidEndSepPunct{\mcitedefaultmidpunct}
{\mcitedefaultendpunct}{\mcitedefaultseppunct}\relax
\EndOfBibitem
\bibitem[S.~Marques \latin{et~al.}(2020)S.~Marques, P.~O.~Nogueira,
  F.~Dillenburg, C.~Barbosa, and Bordin]{Marques20}
S.~Marques,~M.; P.~O.~Nogueira,~T.; F.~Dillenburg,~R.; C.~Barbosa,~M.;
  Bordin,~J.~R. Waterlike anomalies in hard core–soft shell nanoparticles
  using an effective potential approach: Pinned vs adsorbed polymers.
  \emph{Journal of Applied Physics} \textbf{2020}, \emph{127}, 054701\relax
\mciteBstWouldAddEndPuncttrue
\mciteSetBstMidEndSepPunct{\mcitedefaultmidpunct}
{\mcitedefaultendpunct}{\mcitedefaultseppunct}\relax
\EndOfBibitem
\bibitem[Lafitte \latin{et~al.}(2014)Lafitte, Kumar, and
  Panagiotopoulos]{Lafitte14}
Lafitte,~T.; Kumar,~S.~K.; Panagiotopoulos,~A.~Z. Self-assembly of
  polymer-grafted nanoparticles in thin films. \emph{Soft Matter}
  \textbf{2014}, \emph{10}, 786--794\relax
\mciteBstWouldAddEndPuncttrue
\mciteSetBstMidEndSepPunct{\mcitedefaultmidpunct}
{\mcitedefaultendpunct}{\mcitedefaultseppunct}\relax
\EndOfBibitem
\bibitem[Curk \latin{et~al.}(2014)Curk, Martinez-Veracoechea, Frenkel, and
  Dobnikar]{Curk14}
Curk,~T.; Martinez-Veracoechea,~F.~J.; Frenkel,~D.; Dobnikar,~J. Nanoparticle
  Organization in Sandwiched Polymer Brushes. \emph{Nano Letters}
  \textbf{2014}, \emph{14}, 2617--2622\relax
\mciteBstWouldAddEndPuncttrue
\mciteSetBstMidEndSepPunct{\mcitedefaultmidpunct}
{\mcitedefaultendpunct}{\mcitedefaultseppunct}\relax
\EndOfBibitem
\bibitem[Nie \latin{et~al.}(2016)Nie, Li, Wang, and Zhang]{Nie16}
Nie,~G.; Li,~G.; Wang,~L.; Zhang,~X. Nanocomposites of polymer brush and
  inorganic nanoparticles: preparation{,} characterization and application.
  \emph{Polym. Chem.} \textbf{2016}, \emph{7}, 753--769\relax
\mciteBstWouldAddEndPuncttrue
\mciteSetBstMidEndSepPunct{\mcitedefaultmidpunct}
{\mcitedefaultendpunct}{\mcitedefaultseppunct}\relax
\EndOfBibitem
\bibitem[Wang \latin{et~al.}(2016)Wang, Zheng, Liu, Wu, and Zhang]{Wang16}
Wang,~Z.; Zheng,~Z.; Liu,~J.; Wu,~Y.; Zhang,~L. Tuning the Mechanical
  Properties of Polymer Nanocomposites Filled with Grafted Nanoparticles by
  Varying the Grafted Chain Length and Flexibility. \emph{Polymers}
  \textbf{2016}, \emph{8}, 270\relax
\mciteBstWouldAddEndPuncttrue
\mciteSetBstMidEndSepPunct{\mcitedefaultmidpunct}
{\mcitedefaultendpunct}{\mcitedefaultseppunct}\relax
\EndOfBibitem
\bibitem[Bos \latin{et~al.}(2019)Bos, van~der Scheer, Ellenbroek, and
  Sprakel]{Bos19}
Bos,~I.; van~der Scheer,~P.; Ellenbroek,~W.~G.; Sprakel,~J. Two-dimensional
  crystals of star polymers: a tale of tails. \emph{Soft Matter} \textbf{2019},
  \emph{15}, 615--622\relax
\mciteBstWouldAddEndPuncttrue
\mciteSetBstMidEndSepPunct{\mcitedefaultmidpunct}
{\mcitedefaultendpunct}{\mcitedefaultseppunct}\relax
\EndOfBibitem
\bibitem[Jagla(1998)]{Ja98}
Jagla,~E.~A. Phase behavior of a system of particles with core collapse.
  \emph{Phys. Rev. E} \textbf{1998}, \emph{58}, 1478\relax
\mciteBstWouldAddEndPuncttrue
\mciteSetBstMidEndSepPunct{\mcitedefaultmidpunct}
{\mcitedefaultendpunct}{\mcitedefaultseppunct}\relax
\EndOfBibitem
\bibitem[Saija \latin{et~al.}(2009)Saija, Prestipino, and Malescio]{Saija09}
Saija,~F.; Prestipino,~S.; Malescio,~G. Anomalous phase behavior of a
  soft-repulsive potential with a strictly monotonic force. \emph{Physical
  Review E} \textbf{2009}, \emph{80}, 031502\relax
\mciteBstWouldAddEndPuncttrue
\mciteSetBstMidEndSepPunct{\mcitedefaultmidpunct}
{\mcitedefaultendpunct}{\mcitedefaultseppunct}\relax
\EndOfBibitem
\bibitem[Malescio and Saija(2011)Malescio, and Saija]{Malescio11}
Malescio,~G.; Saija,~F. A Criterion for Anomalous Melting in Systems with
  Isotropic Interactions. \emph{The Journal of Physical Chemistry B}
  \textbf{2011}, \emph{115}, 14091--14098\relax
\mciteBstWouldAddEndPuncttrue
\mciteSetBstMidEndSepPunct{\mcitedefaultmidpunct}
{\mcitedefaultendpunct}{\mcitedefaultseppunct}\relax
\EndOfBibitem
\bibitem[Prestipino \latin{et~al.}(2010)Prestipino, Saija, and
  Malescio]{Prestipino10}
Prestipino,~S.; Saija,~F.; Malescio,~G. Anomalous phase behavior in a model
  fluid with only one type of local structure. \emph{Journal of Chemical
  Physics} \textbf{2010}, \emph{133}, 144504\relax
\mciteBstWouldAddEndPuncttrue
\mciteSetBstMidEndSepPunct{\mcitedefaultmidpunct}
{\mcitedefaultendpunct}{\mcitedefaultseppunct}\relax
\EndOfBibitem
\bibitem[Prestipino \latin{et~al.}(2012)Prestipino, Saija, and
  Giaquinta]{Prestipino12}
Prestipino,~S.; Saija,~F.; Giaquinta,~P.~V. Hexatic phase and water-like
  anomalies in a two-dimensional fluid of particles with a weakly softened
  core. \emph{Journal of Chemical Physics} \textbf{2012}, \emph{137},
  104503\relax
\mciteBstWouldAddEndPuncttrue
\mciteSetBstMidEndSepPunct{\mcitedefaultmidpunct}
{\mcitedefaultendpunct}{\mcitedefaultseppunct}\relax
\EndOfBibitem
\bibitem[Coslovich and Ikeda(2013)Coslovich, and Ikeda]{Cos13}
Coslovich,~D.; Ikeda,~A. Cluster and reentrant anomalies of nearly Gaussian
  core particles. \emph{Soft Matter} \textbf{2013}, \emph{9}, 6786\relax
\mciteBstWouldAddEndPuncttrue
\mciteSetBstMidEndSepPunct{\mcitedefaultmidpunct}
{\mcitedefaultendpunct}{\mcitedefaultseppunct}\relax
\EndOfBibitem
\bibitem[Almarza \latin{et~al.}(2016)Almarza, Pekalski, and Ciach]{almarza2016}
Almarza,~N.~G.; Pekalski,~J.; Ciach,~A. Effects of confinement on pattern
  formation in two dimensional systems with competing interactions. \emph{Soft
  Matter} \textbf{2016}, \emph{12}, 7551--7563\relax
\mciteBstWouldAddEndPuncttrue
\mciteSetBstMidEndSepPunct{\mcitedefaultmidpunct}
{\mcitedefaultendpunct}{\mcitedefaultseppunct}\relax
\EndOfBibitem
\bibitem[Litniewski and Ciach(2019)Litniewski, and Ciach]{ciach2019jcp}
Litniewski,~M.; Ciach,~A. Effect of aggregation on adsorption phenomena.
  \emph{The Journal of Chemical Physics} \textbf{2019}, \emph{150},
  234702\relax
\mciteBstWouldAddEndPuncttrue
\mciteSetBstMidEndSepPunct{\mcitedefaultmidpunct}
{\mcitedefaultendpunct}{\mcitedefaultseppunct}\relax
\EndOfBibitem
\bibitem[Pekalski \latin{et~al.}(2019)Pekalski, Bildanau, and
  Ciach]{ciachsoft2019}
Pekalski,~J.; Bildanau,~E.; Ciach,~A. Self-assembly of spiral patterns in
  confined systems with competing interactions. \emph{Soft Matter}
  \textbf{2019}, \emph{15}, 7715--7721\relax
\mciteBstWouldAddEndPuncttrue
\mciteSetBstMidEndSepPunct{\mcitedefaultmidpunct}
{\mcitedefaultendpunct}{\mcitedefaultseppunct}\relax
\EndOfBibitem
\bibitem[Pekalski \latin{et~al.}(2020)Pekalski, Rządkowski, and
  Panagiotopoulos]{grego2020}
Pekalski,~J.; Rządkowski,~W.; Panagiotopoulos,~A.~Z. Shear-induced ordering in
  systems with competing interactions: A machine learning study. \emph{The
  Journal of Chemical Physics} \textbf{2020}, \emph{152}, 204905\relax
\mciteBstWouldAddEndPuncttrue
\mciteSetBstMidEndSepPunct{\mcitedefaultmidpunct}
{\mcitedefaultendpunct}{\mcitedefaultseppunct}\relax
\EndOfBibitem
\bibitem[Bildanau \latin{et~al.}(2020)Bildanau, P\ifmmode~\mbox{\k{e}}\else
  \k{e}\fi{}kalski, Vikhrenko, and Ciach]{ciach2020}
Bildanau,~E.; P\ifmmode~\mbox{\k{e}}\else \k{e}\fi{}kalski,~J.; Vikhrenko,~V.;
  Ciach,~A. Adsorption anomalies in a two-dimensional model of cluster-forming
  systems. \emph{Phys. Rev. E} \textbf{2020}, \emph{101}, 012801\relax
\mciteBstWouldAddEndPuncttrue
\mciteSetBstMidEndSepPunct{\mcitedefaultmidpunct}
{\mcitedefaultendpunct}{\mcitedefaultseppunct}\relax
\EndOfBibitem
\bibitem[Salcedo \latin{et~al.}(2011)Salcedo, de~Oliveira, Barraz, Chakravarty,
  and Barbosa]{Evy2011}
Salcedo,~E.; de~Oliveira,~A.~B.; Barraz,~N.~M.; Chakravarty,~C.; Barbosa,~M.~C.
  Core-softened fluids, water-like anomalies, and the liquid-liquid critical
  points. \emph{The Journal of Chemical Physics} \textbf{2011}, \emph{135},
  044517\relax
\mciteBstWouldAddEndPuncttrue
\mciteSetBstMidEndSepPunct{\mcitedefaultmidpunct}
{\mcitedefaultendpunct}{\mcitedefaultseppunct}\relax
\EndOfBibitem
\bibitem[Fomin \latin{et~al.}(2011)Fomin, Tsiok, and Ryzhov]{Formin2011}
Fomin,~Y.~D.; Tsiok,~E.~N.; Ryzhov,~V.~N. Inversion of sequence of diffusion
  and density anomalies in core-softened systems. \emph{The Journal of Chemical
  Physics} \textbf{2011}, \emph{135}, 234502\relax
\mciteBstWouldAddEndPuncttrue
\mciteSetBstMidEndSepPunct{\mcitedefaultmidpunct}
{\mcitedefaultendpunct}{\mcitedefaultseppunct}\relax
\EndOfBibitem
\bibitem[Barraz \latin{et~al.}(2009)Barraz, Salcedo, and Barbosa]{Ney2009}
Barraz,~N.~M.; Salcedo,~E.; Barbosa,~M.~C. Thermodynamic, dynamic, and
  structural anomalies for shoulderlike potentials. \emph{The Journal of
  Chemical Physics} \textbf{2009}, \emph{131}, 094504\relax
\mciteBstWouldAddEndPuncttrue
\mciteSetBstMidEndSepPunct{\mcitedefaultmidpunct}
{\mcitedefaultendpunct}{\mcitedefaultseppunct}\relax
\EndOfBibitem
\bibitem[da~Silva \latin{et~al.}(2010)da~Silva, Salcedo, de~Oliveira, and
  Barbosa]{jonatan2010}
da~Silva,~J.~N.; Salcedo,~E.; de~Oliveira,~A.~B.; Barbosa,~M.~C. Effects of the
  attractive interactions in the thermodynamic, dynamic, and structural
  anomalies of a two length scale potential. \emph{The Journal of Chemical
  Physics} \textbf{2010}, \emph{133}, 244506\relax
\mciteBstWouldAddEndPuncttrue
\mciteSetBstMidEndSepPunct{\mcitedefaultmidpunct}
{\mcitedefaultendpunct}{\mcitedefaultseppunct}\relax
\EndOfBibitem
\bibitem[Limbach \latin{et~al.}(2006)Limbach, Arnold, Mann, and
  Holm]{espresso1}
Limbach,~H.-J.; Arnold,~A.; Mann,~B.~A.; Holm,~C. ESPResSo - An Extensible
  Simulation Package for Research on Soft Matter Systems. \emph{Comput. Phys.
  Commun.} \textbf{2006}, \emph{174}, 704--727\relax
\mciteBstWouldAddEndPuncttrue
\mciteSetBstMidEndSepPunct{\mcitedefaultmidpunct}
{\mcitedefaultendpunct}{\mcitedefaultseppunct}\relax
\EndOfBibitem
\bibitem[Arnold \latin{et~al.}(2013)Arnold, Lenz, Kesselheim, Weeber,
  Fahrenberger, Roehm, Kosovan, and Holm]{espresso2}
Arnold,~A.; Lenz,~O.; Kesselheim,~S.; Weeber,~R.; Fahrenberger,~F.; Roehm,~D.;
  Kosovan,~P.; Holm,~C. In \emph{Meshfree Methods for Partial Differential
  Equations VI}; Griebel,~M., Schweitzer,~M.~A., Eds.; Lecture Notes in
  Computational Science and Engineering; Springer Berlin Heidelberg, 2013;
  Vol.~89; pp 1--23\relax
\mciteBstWouldAddEndPuncttrue
\mciteSetBstMidEndSepPunct{\mcitedefaultmidpunct}
{\mcitedefaultendpunct}{\mcitedefaultseppunct}\relax
\EndOfBibitem
\bibitem[Bordin \latin{et~al.}(2012)Bordin, Diehl, Barbosa, and
  Levin]{Bordin12}
Bordin,~J.~R.; Diehl,~A.; Barbosa,~M.~C.; Levin,~Y. Ion fluxes through
  nanopores and transmembrane channels. \emph{Phys. Rev. E} \textbf{2012},
  \emph{85}, 031914\relax
\mciteBstWouldAddEndPuncttrue
\mciteSetBstMidEndSepPunct{\mcitedefaultmidpunct}
{\mcitedefaultendpunct}{\mcitedefaultseppunct}\relax
\EndOfBibitem
\bibitem[Bordin \latin{et~al.}(2013)Bordin, Diehl, and Barbosa]{Bordin13}
Bordin,~J.~R.; Diehl,~A.; Barbosa,~M.~C. Relation Between Flow Enhancement
  Factor and Structure for Core-Softened Fluids Inside Nanotubes. \emph{The
  Journal of Physical Chemistry B} \textbf{2013}, \emph{117}, 7047--7056, PMID:
  23692639\relax
\mciteBstWouldAddEndPuncttrue
\mciteSetBstMidEndSepPunct{\mcitedefaultmidpunct}
{\mcitedefaultendpunct}{\mcitedefaultseppunct}\relax
\EndOfBibitem
\bibitem[Bordin \latin{et~al.}(2014)Bordin, Andrade, Diehl, and
  Barbosa]{Bordin14}
Bordin,~J.~R.; Andrade,~J.~S.; Diehl,~A.; Barbosa,~M.~C. Enhanced flow of
  core-softened fluids through narrow nanotubes. \emph{The Journal of Chemical
  Physics} \textbf{2014}, \emph{140}, 194504\relax
\mciteBstWouldAddEndPuncttrue
\mciteSetBstMidEndSepPunct{\mcitedefaultmidpunct}
{\mcitedefaultendpunct}{\mcitedefaultseppunct}\relax
\EndOfBibitem
\bibitem[Allen and Tildesley(2017)Allen, and Tildesley]{allen2017}
Allen,~M.; Tildesley,~D. \emph{Computer Simulation of Liquids: Second Edition};
  OUP Oxford, 2017\relax
\mciteBstWouldAddEndPuncttrue
\mciteSetBstMidEndSepPunct{\mcitedefaultmidpunct}
{\mcitedefaultendpunct}{\mcitedefaultseppunct}\relax
\EndOfBibitem
\bibitem[Krott \latin{et~al.}(2015)Krott, Bordin, and Barbosa]{Bordin15}
Krott,~L.~B.; Bordin,~J.~R.; Barbosa,~M.~C. New Structural Anomaly Induced by
  Nanoconfinement. \emph{The Journal of Physical Chemistry B} \textbf{2015},
  \emph{119}, 291--300, PMID: 25494049\relax
\mciteBstWouldAddEndPuncttrue
\mciteSetBstMidEndSepPunct{\mcitedefaultmidpunct}
{\mcitedefaultendpunct}{\mcitedefaultseppunct}\relax
\EndOfBibitem
\bibitem[Toledano \latin{et~al.}(2009)Toledano, Sciortino, and
  Zaccarelli]{Toledano09}
Toledano,~J. C.~F.; Sciortino,~F.; Zaccarelli,~E. Colloidal systems with
  competing interactions: from an arrested repulsive cluster phase to a gel.
  \emph{Soft Matter} \textbf{2009}, \emph{5}, 2390--2398\relax
\mciteBstWouldAddEndPuncttrue
\mciteSetBstMidEndSepPunct{\mcitedefaultmidpunct}
{\mcitedefaultendpunct}{\mcitedefaultseppunct}\relax
\EndOfBibitem
\bibitem[{Rafael Bordin}(2019)]{Bordin19}
{Rafael Bordin},~J. Distinct self-assembly aggregation patters of nanorods with
  decorated ends: A simple model study. \emph{Fluid Phase Equilibria}
  \textbf{2019}, \emph{499}, 112251\relax
\mciteBstWouldAddEndPuncttrue
\mciteSetBstMidEndSepPunct{\mcitedefaultmidpunct}
{\mcitedefaultendpunct}{\mcitedefaultseppunct}\relax
\EndOfBibitem
\bibitem[Bordin(2018)]{Bordin18b}
Bordin,~J.~R. Distinct aggregation patterns and fluid porous phase in a 2D
  model for colloids with competitive interactions. \emph{Physica A:
  Statistical Mechanics and its Applications} \textbf{2018}, \emph{495}, 215 --
  224\relax
\mciteBstWouldAddEndPuncttrue
\mciteSetBstMidEndSepPunct{\mcitedefaultmidpunct}
{\mcitedefaultendpunct}{\mcitedefaultseppunct}\relax
\EndOfBibitem
\bibitem[Dudalov \latin{et~al.}(2014)Dudalov, Fomin, Tsiok, and Ryzhov]{Duda14}
Dudalov,~D.~E.; Fomin,~Y.~D.; Tsiok,~E.~N.; Ryzhov,~V.~N. How dimensionality
  changes the anomalous behavior and melting scenario of a core-softened
  potential system? \emph{Soft Matter} \textbf{2014}, \emph{10}, 4966\relax
\mciteBstWouldAddEndPuncttrue
\mciteSetBstMidEndSepPunct{\mcitedefaultmidpunct}
{\mcitedefaultendpunct}{\mcitedefaultseppunct}\relax
\EndOfBibitem
\bibitem[Fomin \latin{et~al.}(2020)Fomin, Ryzhov, and Tsiok]{Fomin20}
Fomin,~Y.~D.; Ryzhov,~V.~N.; Tsiok,~E.~N. Interplay between freezing and
  density anomaly in a confined core-softened fluid. \emph{Molecular Physics}
  \textbf{2020}, \emph{118}, e1718792\relax
\mciteBstWouldAddEndPuncttrue
\mciteSetBstMidEndSepPunct{\mcitedefaultmidpunct}
{\mcitedefaultendpunct}{\mcitedefaultseppunct}\relax
\EndOfBibitem
\bibitem[Bordin and Barbosa(2018)Bordin, and Barbosa]{Bordin18}
Bordin,~J.~R.; Barbosa,~M.~C. Waterlike anomalies in a two-dimensional
  core-softened potential. \emph{Phys. Rev. E} \textbf{2018}, \emph{97},
  022604\relax
\mciteBstWouldAddEndPuncttrue
\mciteSetBstMidEndSepPunct{\mcitedefaultmidpunct}
{\mcitedefaultendpunct}{\mcitedefaultseppunct}\relax
\EndOfBibitem
\bibitem[Krott and Bordin(2013)Krott, and Bordin]{Krott13b}
Krott,~L.; Bordin,~J.~R. Distinct dynamical and structural properties of a
  core-softened sfluid when confined between fluctuating and fixed walls.
  \emph{Journal of Chemical Physics} \textbf{2013}, \emph{139}, 154502\relax
\mciteBstWouldAddEndPuncttrue
\mciteSetBstMidEndSepPunct{\mcitedefaultmidpunct}
{\mcitedefaultendpunct}{\mcitedefaultseppunct}\relax
\EndOfBibitem
\bibitem[Köhler \latin{et~al.}(2019)Köhler, Bordin, {de Matos}, and
  Barbosa]{Kohler19}
Köhler,~M.~H.; Bordin,~J.~R.; {de Matos},~C.~F.; Barbosa,~M.~C. Water in
  nanotubes: The surface effect. \emph{Chemical Engineering Science}
  \textbf{2019}, \emph{203}, 54 -- 67\relax
\mciteBstWouldAddEndPuncttrue
\mciteSetBstMidEndSepPunct{\mcitedefaultmidpunct}
{\mcitedefaultendpunct}{\mcitedefaultseppunct}\relax
\EndOfBibitem
\bibitem[Köhler \latin{et~al.}(2018)Köhler, Bordin, {da Silva}, and
  Barbosa]{Kohler18}
Köhler,~M.~H.; Bordin,~J.~R.; {da Silva},~L.~B.; Barbosa,~M.~C. Structure and
  dynamics of water inside hydrophobic and hydrophilic nanotubes. \emph{Physica
  A: Statistical Mechanics and its Applications} \textbf{2018}, \emph{490}, 331
  -- 337\relax
\mciteBstWouldAddEndPuncttrue
\mciteSetBstMidEndSepPunct{\mcitedefaultmidpunct}
{\mcitedefaultendpunct}{\mcitedefaultseppunct}\relax
\EndOfBibitem
\bibitem[Bordin and Barbosa(2017)Bordin, and Barbosa]{Bordin17}
Bordin,~J.~R.; Barbosa,~M.~C. Flow and structure of fluids in functionalized
  nanopores. \emph{Physica A: Statistical Mechanics and its Applications}
  \textbf{2017}, \emph{467}, 137 -- 147\relax
\mciteBstWouldAddEndPuncttrue
\mciteSetBstMidEndSepPunct{\mcitedefaultmidpunct}
{\mcitedefaultendpunct}{\mcitedefaultseppunct}\relax
\EndOfBibitem
\bibitem[Nogueira \latin{et~al.}(2020)Nogueira, Frota, Piazza, and
  Bordin]{Nogueira20}
Nogueira,~T. P.~O.; Frota,~H.~O.; Piazza,~F.; Bordin,~J.~R. Tracer diffusion in
  crowded solutions of sticky polymers. \emph{Phys. Rev. E} \textbf{2020},
  \emph{102}, 032618\relax
\mciteBstWouldAddEndPuncttrue
\mciteSetBstMidEndSepPunct{\mcitedefaultmidpunct}
{\mcitedefaultendpunct}{\mcitedefaultseppunct}\relax
\EndOfBibitem
\bibitem[Bordin \latin{et~al.}(2012)Bordin, de~Oliveira, Diehl, and
  Barbosa]{Bordin12b}
Bordin,~J.~R.; de~Oliveira,~A.~B.; Diehl,~A.; Barbosa,~M.~C. Diffusion
  enhancement in core-softened fluid confined in nanotubes. \emph{The Journal
  of Chemical Physics} \textbf{2012}, \emph{137}, 084504\relax
\mciteBstWouldAddEndPuncttrue
\mciteSetBstMidEndSepPunct{\mcitedefaultmidpunct}
{\mcitedefaultendpunct}{\mcitedefaultseppunct}\relax
\EndOfBibitem
\end{mcitethebibliography}

\providecommand{\latin}[1]{#1}
\makeatletter
\providecommand{\doi}
  {\begingroup\let\do\@makeother\dospecials
  \catcode`\{=1 \catcode`\}=2 \doi@aux}
\providecommand{\doi@aux}[1]{\endgroup\texttt{#1}}
\makeatother
\providecommand*\mcitethebibliography{\thebibliography}
\csname @ifundefined\endcsname{endmcitethebibliography}
  {\let\endmcitethebibliography\endthebibliography}{}

\end{document}